\documentclass[leqno]{article}
\usepackage{amssymb,latexsym,amsmath,mathcomp}

\usepackage{graphicx}	
\usepackage{amssymb}	

\usepackage{natbib}      

\usepackage{mathcomp}
\usepackage[]{endfloat} 
\usepackage{siunitx}     
\usepackage[version=3]{mhchem} 
\usepackage{longtable} 

\newcommand{\epsCG}{  \epsilon^{\scriptscriptstyle{C}}_{\scriptscriptstyle{G}}  }
\newcommand{\tcmuCG}{  \tcmu^{\scriptscriptstyle{C}}_{\scriptscriptstyle{G}}  }

\newcommand{\boxemptyplusCG}{  \square^{\scriptscriptstyle{C^{+}}}_{\scriptscriptstyle{G}}  }
\newcommand{\boxemptyminusCG}{  \square^{\scriptscriptstyle{C^{-}}}_{\scriptscriptstyle{G}}  }

\newcommand{\Msun}{  M_{\odot}  }
\newcommand{\Qsun}{  Q_{\odot}^{+}  }
\newcommand{\Rsun}{  R_{\odot}  }

\newcommand{\tcmuboxdotplus}{  \tcmu^{\scriptscriptstyle{+}}_{\scriptstyle{\square}}  }
\newcommand{\tcmuboxdotminus}{  \tcmu^{\scriptscriptstyle{-}}_{\scriptstyle{\square}}  }
\newcommand{\tcmuodotplus}{  \tcmu^{\scriptscriptstyle{+}}_{\scriptscriptstyle{\odot}}  }
\newcommand{\tcmuboxplusplusplus}{  \tcmu^{\scriptscriptstyle{++}}_{\scriptscriptstyle{\boxplus}}  }

\newcommand{\codot} { {c}^{}_{\scriptstyle{\square }}   }
\newcommand{\cboxplus} { {c}^{}_{\scriptscriptstyle{\boxplus }}   }
\newcommand{\cboxminus}{ c^{}_{\scriptscriptstyle{\boxminus}}  }

\newcommand{\fboxdot}{ f^{}_{\scriptstyle{\Box}}  }

\newcommand{\tcmuboxplus}{  \tcmu^{}_{\scriptscriptstyle{\boxplus   }}  }
\newcommand{\tcmuboxminus}{  \tcmu^{}_{\scriptscriptstyle{\boxminus}}  }
\newcommand{\tcmuboxdot}{  \tcmu^{}_{\scriptstyle{\square}}  }

\newcommand{\rhoplus}{  \rho^{}_{\scriptscriptstyle{\boxplus   }}  }
\newcommand{\rhodot}{  \rho^{}_{\scriptstyle{\square}}  }
\newcommand{\moverbarplus}{  \overline{m}^{}_{\scriptscriptstyle{\boxplus   }}  }
\newcommand{\moverbardot}{  \overline{m}^{}_{{\square}}  }

\newcommand{\cGBH}{  c^{}_{\scriptscriptstyle{GBH}}   }
\newcommand{\cBH}{  c^{}_{\scriptscriptstyle{BH}}   }
\newcommand{\MGBH}{  M^{}_{\scriptscriptstyle{GBH}}   }
\newcommand{\Mboxdot}{ M^{}_{\scriptstyle{\square}}  }
\newcommand{\Mboxplus}{ M^{}_{\scriptscriptstyle{\boxplus}}  }
\newcommand{\ML}{ M^{}_{\scriptscriptstyle{L}}  }
\newcommand{\RL}{ R^{}_{\scriptscriptstyle{L}}  }
\newcommand{\Qboxplus}{ Q^{}_{\scriptscriptstyle{\boxplus}}  }

\newcommand{\csplus}{ c^{}_{S\scriptscriptstyle{\boxplus}}  }
\newcommand{\csplussquared}{ c^{2}_{S\scriptscriptstyle{\boxplus}}  }

\newcommand{\cboxplussquared} { {c}^{2}_{\scriptscriptstyle{\boxplus }}   }

\begin{document}

\title{Cosmological Implications of Trace-Charged Dark Matter}
\author{Jason P. Morgan \thanks{jason.morgan@rhul.ac.uk}\\
 Earth Science Department\\
 Royal Holloway, University of London\\
 Runnymede Borough, Surrey, TW20 0EX UK}
\maketitle

\begin{abstract}
Trace charge imbalances can explain puzzling cosmological observations
such as the large `missing' fraction of electrons in cosmic rays and
their contrast to the charge-neutral solar wind, the extreme energy
sources that sustain pulsars, quasars, galactic jets and active galactic nuclei (AGN), 
the origin and nature of `dark matter' galaxy haloes, and the 
apparent acceleration of the expansion of the Universe. When there are 
$\sim \num{9e-19}$ amounts of excess $\ce{H3+}$ or $\ce{H-}$ within 
cold diffuse clouds of $\ce{H2}$, residual repulsive Coulomb 
forces between these few unbalanced charges are comparable to the gravitational attractions between the many nucleons. 
Thus, trace-charged dark matter is inert with respect 
to static electrogravitional self-attractions, but responds 
to electromagnetic fields and gravitational attractions with uncharged 
matter. Trace charge is also the ionic catalyst 
that keeps dark matter in the state of unseen clouds 
of cold molecular hydrogen plus trace $\ce{H3+}$. Once warm enough to partially 
ionize, bright matter preferentially expels its net charge to 
become nearly charge-neutral. Planets surrounding stars become charge neutral as they 
bathe in a charge-neutral stellar wind. In contrast, around trace-charged AGNs, black holes, and pulsars,
newly ionized protons along with a significant fraction of entrained 
dark matter are Coulomb-expelled to relativistic velocities in polar 
jets. Extrasolar cosmic rays generated by these would be 
strongly proton-dominated, as observed. Initial trace charge imbalances 
could originate at the onset of the the Big Bang by segregation of an electron-rich shell to peripheral
expanding matter of the Universe.
\end{abstract}





\section{Introduction}

Matter dominates over antimatter in our local universe. Several lines of evidence suggest that the region illuminated by the Big Bang (e.g. the ``Universe") is also strongly matter dominated. In spite of the $\sim 1836$ times   larger mass of the positive charge carrying proton than its negative charge carrying electron counterpart, it has been almost an axiom of cosmology that our local Universe has a local net charge of zero, even though explosive ionized events should lead to temporary mass-dependent charge segregation between faster less-massive electrons and slower moving protons. Only in general relativity does there appear to have been previous discussion of scenarios in which trace-charge imbalances play a key role in resisting gravitational attraction (\cite{weyl1917,maju1947,bonn1980}), but its cosmological implications -- with the important exception of gravitational red-shifts (\cite{Bonn1975,bond1999}) -- appear to remain unexplored.

Here I will explore some consequences of a scenario in which there exist net local charge imbalances at the level where these imbalances generate electrostatic Coulomb forces of similar size to gravitational attractions, associated with a net charge-to-mass ratio of order $\epsCG =\num{8.6167e-11} \si{C/kg}$, or, alternatively expressed, a net proton or electron imbalance of 1 extra electron/proton per the mass equivalent of $\num{5.56e17}$ hydrogen molecules.  (See Table 1 for definitions of $\epsCG$ and the other symbols used here. A $\boxplus$ will be used to denote trace positively charged `dark' matter, a $\boxminus$ to denote trace negatively charged `dark' matter, and a $\square$ to denote nearly uncharged warm and bright matter.) This hypothesis will be shown to lead to simple conventional physics explanations for effects commonly attributed to unconventional forms of dark energy or dark matter, often called $\Lambda CDM$.  For example, the strong jetting observed in AGNs can be created if matter ionizes near the event horizon of a positively trace-charged black hole. Newly ionized protons feel a strong Coulomb repulsion from the black hole, and are jetted away from its magnetic poles, thereby creating a proton-rich source for the  strong charge imbalance seen in cosmic rays. Intense magnetic fields associated with pulsars and neutron stars can be explained by the neutron star being trace charged and rapidly rotating as a byproduct of its supernova-linked birth. The dark matter clouds forming `Galactic Halos' can be explained as being clouds of hard-to-see cold molecular hydrogen containing a small net positive charge in the form of hard-to-see cold \ce{H3+}. These abundant unseen `dark matter' clouds are mixed with radio-visible HI-clouds of cold atomic hydrogen \ce{H1},  providing the unseen mass needed to explain the `too-fast' rotation of radiovisible HI-clouds.  These and other predictions will be quantitatively assessed below. 

This hypothesis also implies that stellar fusion generates bright matter $\tcmuboxdot$ in stars and the planets orbiting within their heliospheres that is of order $\num{9e-19} ( \simeq \cboxplus ( \tcmuboxdotplus ))$ to $\num{4.9e-22} (\simeq \cboxminus ( \tcmuboxdotminus ) )$ \textit{less} charged than its surrounding cloud of positively trace-charged dark matter $\tcmuboxplus$ or negatively trace-charged dark matter $\tcmuboxminus$,  of order one extra proton per $\num{1.2e36}$ hydrogen atoms or one extra electron  per $\num{2.3e39}$ hydrogen atoms.   Black holes, in contrast, would typically accrete some trace-charged dark matter and retain their net charge because they cannot expel the excited ions associated with fusion reactions. Neutron stars would also tend to form containing trace-charged matter that they retain as they evolve. The observation that cosmic rays have $\sim 50$ positively charged particles for each electron implies that nearby dark matter has a trace positive charge. I will explore the implications of this for the generation of stellar winds from bright matter and the generation of cosmic rays by axial jetting from accreting black holes, active galactic nuclei, and neutron stars. Then I will explore the interpretation of galactic dark matter halos within this conceptual framework.
In a final, speculative, section of this paper I will explore a supernova-like Big Bang scenario in which these local charge imbalances appear to naturally arise.

\section{The physics and chemistry of trace-charged dark matter}

In trace-charged dark matter there is an approximate balance between Coulomb repulsive forces and gravitational attraction. A standard assumption in cosmology is that the Coulomb attraction between electrons and protons is so strong, relative to gravitational forces, that perfect charge equilibrium will spontaneously reoccur. But what if charge reequilibrates only to a level where residual Coulomb forces are comparable to the gravitational forces between hadrons? It only takes a $\num{9e-19}$ imbalance between protons and electrons to induce a repulsive Coulomb force equal to the gravitational attraction of this slightly charged proton+electon mixture. Equivalently, a charge/mass ratio 
$\epsCG =\num{8.6167e-11} \si{C/kg}$ leads to its Coulomb repulsion (proportional to $k_E(\epsCG)^{2}m^2$, with Coulomb's constant $k_E=\num{8.987551787e9} \si{N.m^{2}.C^{-2}}$) being as strong as its gravitational attraction (proportional to $Gm^{2}$, with $G=\num{6.673e-11} \si{N.m^{2}.kg^{-2}}$). Trace-charged dark matter $\tcmuboxplus$ or $\tcmuboxminus$ with $|\tcmuCG| =\epsCG$ will be electrogravitationally neutral, $\tcmuboxplus$ or $\tcmuboxminus$ with $|\tcmuCG|<\epsCG$ will  self-attract, and with $|\tcmuCG|>\epsCG$ will self-repel. In the following, I will explore scenarios in which the Universe contains abundant trace-charged dark matter that is nearly electrogravitationally neutral or faintly repulsive. While being almost electrogravitationally neutral, this form of dark matter will still be gravitationally attracted to nearly chargeless warm and bright matter. Due to its net trace charge, it will also be influenced by electromagnetic fields.

In cold matter, free ions are chemically unstable. Instead, their preferred lowest energy state is to be bound to atoms by molecular binding forces, e.g. chemical effects. The chemistry of hydrogen plays a key role in the initial state of both  $\tcmuboxminus$ and $\tcmuboxplus$. Hydrogen has the highest electron affinity (\cite{Rien2002}) of the elements formed by the Big Bang (hydrogen, helium, lithium, and beryllium), therefore \ce{H-} (i.e.\ $1p^{+}2e^{-}$, binding energy or electron affinity $-0.75eV$) was the initial likely negative trace-charge carrier for $\tcmuboxminus$. $\ce{C}$ (electron affinity $-1.26eV$, atomic abundance $5x10^{-3}$) and $\ce{O}$ (electron affinity $-1.46eV$, atomic abundance $10^{-2}$) are fairly abundant elements created by later stellar nucleosynthesis that have a higher electron affinity (\cite{Rien2002}) than $\ce{H}$, while $\ce{F}$ (electron affinity $-3.40eV$, atomic abundance $4x10^{-7}$) and $\ce{Cl}$ (electron affinity $-3.61eV$, atomic abundance $10^{-6}$) are the elements with the highest electron affinities to which free electrons would preferentially migrate over time. 

The positive trace-charge carrier in $\tcmuboxplus$ is likely to also be hydrogen, but in the form of almost unseeable cold \ce{H3+} (i.e.\ $3p^{+}2e^{-}$, binding energy $-4.38eV$) (\cite{cosb1988}) instead of \ce{H2+} (i.e.\ $2p^{+}1e^{-}$, binding energy $-2.65eV$) 
(\cite{zhan2004}). 

Chemical effects will cause most of the cold hydrogen arising from the Big Bang to be in the more stable chemical state of nearly invisible cold molecular hydrogen \ce{H2} (molecular binding energy $-4.78eV$ (\cite{zhan2004})) in preference to atomic hydrogen \ce{H1}, with  trace \ce{H3+} (or \ce{H-}) also serving as the ionic catalysts that greatly speed up low temperature $\ce{H1} + \ce{H1} \to \ce{H2}$ reactions.  \ce{He}, with its high ionization energy $-24.6eV$ and zero electron affinity, is chemically inert in spite of its high atomic abundance.

In the scenarios explored below, most dark matter in our region of the Universe is $\tcmuboxplus$, either in the form of trace-charged black holes, or in the form of diffuse molecular clouds of almost unseeable \ce{H2} and \ce{He} containing about one excess \ce{H3+} molecule  per $\num{5.56e17}$ \ce{H2} (mass-equivalent) molecules, or a $ \ce{H3+}/\ce{H2}$ ratio of $\num{1.8e-18}$.  In other words, $\tcmuboxplus$ is completely `normal' cold hadronic matter, just matter that is slightly spiked with a trace amount of locally charge-unbalanced \ce{H3+}. Note that orders of magnitude larger $10^{-8}-10^{-7}$ abundances of $\ce{H3+}/[\ce{H1} +\ce{H2}]$ have been observed in warmer (therefore radiating and more observable) environments within the Milky Way (\cite{mcca1999,Oka2006,oka1992}), thus the \ce{H3+}  molecule, too, is a well-known molecule of the Universe. Nearby dark matter  $\tcmuboxplus$ may be hiding in plain sight in the Galaxy.

\section{The physics of bright matter}

Once gravitational collapse and star forming occurs in a region where $|\tcmuCG| < \epsCG$, then any ion-emitting warm or bright matter will tend to lose almost all of its trace-excess of net charge. This happens because the charge-to-mass ratio of  a proton  $(\num{9.579e7} \si{C/kg})$ or an electron $(\num{1.759e11} \si{C/kg})$ is enormous with respect to $\epsCG$. For example, if a star ignites with a trace net positive charge, any protons in its stellar wind would feel a net repulsion from the star until the star's charge-to-mass ratio was $ \boxemptyplusCG$ $ \equiv \tcmuodotplus= \num{7.753e-29} \si{C/kg}$ (see Eq. 4 below), while a free electron near a star with a trace net negative charge would feel a net repulsion until the star's net charge-to-mass ratio was $1836$ times smaller, e.g. $\boxemptyminusCG$ $ = \num{-4.222e-32} \si{C/kg}$. While being almost vanishingly close to being charge neutral, such a tiny net charge imbalance in the Sun's bright matter would still have important implications for the generation of the solar wind.

\section{Trace-charged bright matter interpretation of the solar wind}
Lets look in more detail at the electrogravitational binding of protons and electrons in the corona of a positively trace-charged Sun of mass $\Msun =\num{1.989e30}\si{kg}$ and net charge $\Qsun$. The escape velocity $w_{p}$ of a proton is described by the kinetic energy needed to overcome the gravitational and Coulomb potential energy within the solar chromosphere-to-corona  transition region at  radius $\Rsun \simeq \num{7e8}\si{m}$:

\begin{equation}\label{E:Eq1}
\frac{1}{2} m_{p} w_{p}^{2} = \frac{G \Msun m_{p}}  {\Rsun} - \frac{k_{E}\Qsun |e^{-}|}  {\Rsun}
\end{equation}

\noindent
where $|e^{-}|=\num{1.602e-19}\si{C}$ is the elementary charge unit of an electron/proton. This, and the corresponding equation for the escape velocity $w_{e}$ of an electron lead to:

\begin{align}\label{E:Eq2}
 w_{p} &=\sqrt{  \frac{2G \Msun }{\Rsun} - \frac{2k_{E}\Qsun |e^{-}|} {m_{p}\Rsun}   } \\
 w_{e} &=\sqrt{  \frac{2G \Msun }{\Rsun} + \frac{2k_{E}\Qsun |e^{-}|} {m_{e}\Rsun}  } 
\end{align}

\noindent
where the electron mass $m_{e}$ is $\SI{9.109e-31}{kg}$ and the proton mass $m_{e}$ is $\SI{1.673e-27}{kg}$. The maximum or critical charge/mass ratio  $\tcmuodotplus ={\Qsun}/{\Msun}$ without spontaneous proton escape is that which leads to a zero (e.g. non-imaginary) proton escape velocity:

\begin{equation}\label{E:Eq4}
\tcmuodotplus =\frac{Gm_{p}}{k_{E}|e^{-}|} = \num{7.7529e-29} \si{C/kg}
\end{equation}

\begin{figure}
	\includegraphics[width=\columnwidth]{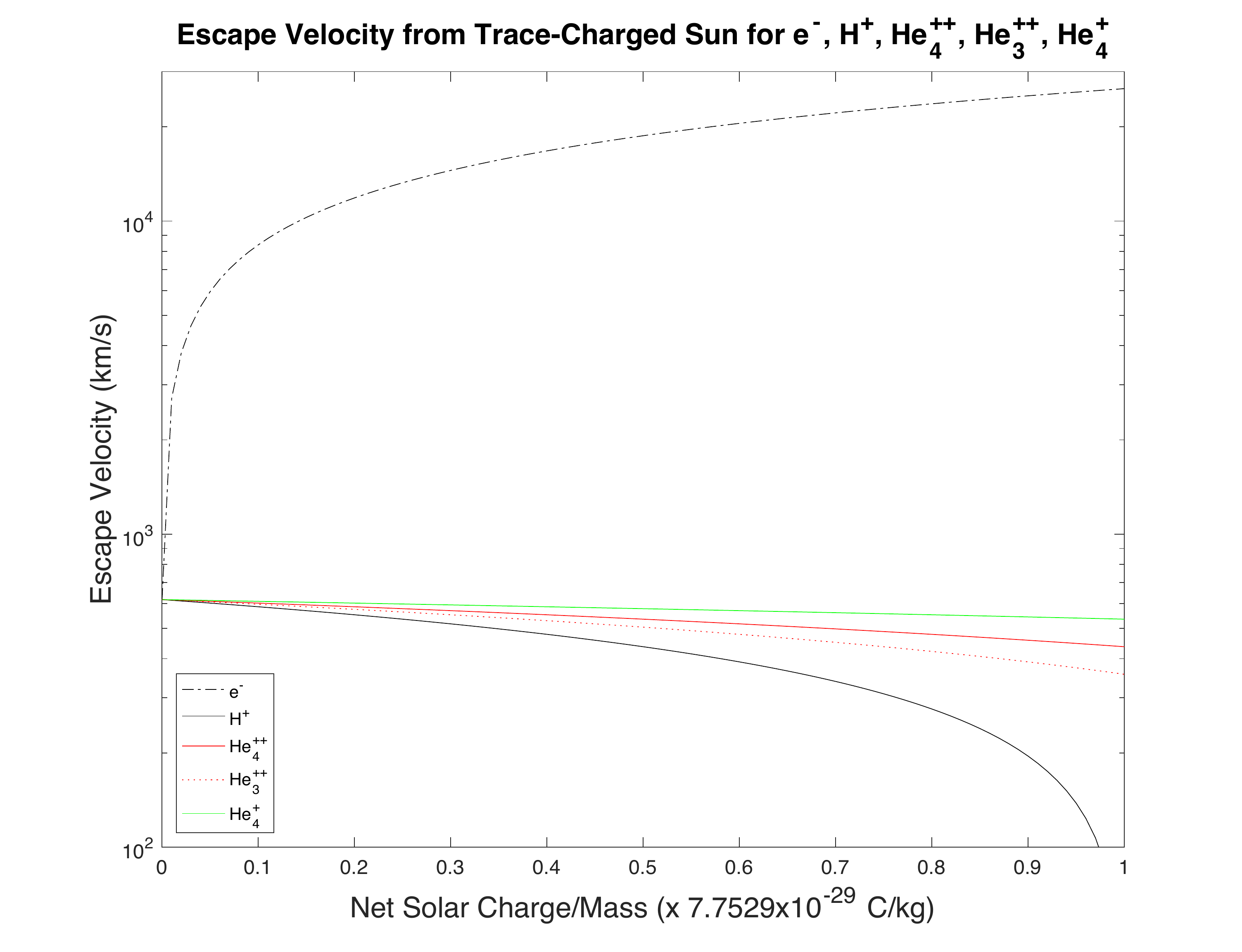}
         \caption{
Escape velocities from a trace-charged Sun of $e^{-}$, $\ce{^{1}H+}$,$\ce{^{4}He++}=\ce{^{2}H+}$, $\ce{^{3}He++}$, and $\ce{^{4}He+}$  as a function of the net solar charge/mass ratio. This ratio is shown as a fraction of the critical charge/mass ratio $\tcmuodotplus = \SI{7.753e-29}{Ckg^{-1}}$. If the Sun's charge/mass ratio is $1/2$ the critical ratio, the trace-charged Sun's escape velocity for protons would be $\sqrt{1/2}$ its gravitational escape velocity of 617 kms$^{-1}$, 436 kms$^{-1}$. In contrast, the escape velocity for electrons would be much higher, $\sim 18,700$ kms$^{-1}$.
}
    \label{fig:Fig1SolarEscapeVelocity} 

\end{figure}
\noindent
Figure 1 shows the charge/mass dependence of the solar escape velocities for free electrons and \ce{^{1}H+}, \ce{^{2}H+} $\equiv$ \ce{^{4}He++}, \ce{^{3}He++}, and  \ce{^{4}He+} ions as the ratio of the net solar charge/mass to $\tcmuodotplus$. Let us make the conventional kinematic theory assumption that that the solar wind forms by a Jeans-like thermal escape process in which a small fraction of electrons and protons have velocities greater than their escape velocity. As long as the Sun's net charge/mass ratio induces a greater flux of escaping positive ions than electrons in the solar wind, the Sun's net charge/mass ratio will decrease. Once the Sun's charge/mass ratio leads to equal escaping fluxes of negative and positive ions, then the ratio will stabilize unless the solar wind's source temperature changes during stellar evolution. 

Solar researchers have explored two candidate velocity distributions to describe the  solar wind 
(\cite{maks1997,meye2007,Pier2010}). The first is the Maxwell-Boltzmann (M-B) velocity distribution $f(v)$ that has an exponential decay distribution within its fast velocity tail

\begin{equation}\label{E:Eq5}
f(v) =\frac{4}{\sqrt{\pi}} \frac{n_{0}}{w_{i}} v^{2} \exp(-v^{2}/w_{i}^{2})
\end{equation}

\noindent
where $w_{i}=\sqrt{{2k_{B}T}/m_{i}}$ is the most probable thermal velocity of a particle of mass $m_{i}$. For the M-B distribution, the escape flux fraction $F(w_{e})$ of particles with velocities greater than the escape velocity $v_{e}$ is given by (\cite{Pier2010})

\begin{equation}\label{E:Eq6}
F(w_{e}) =\frac{n_{0}w_{i}(1+w_{e}^{2}/w_{i}^{2})}{2\sqrt{\pi}}\exp(-w_{e}^{2}/w_{i}^{2})
\end{equation}

\noindent
The second is a `kappa' (also called Lorentzian) distribution (\cite{Pier2010}) with $\kappa \sim 500$ (e.g. M-B-like) describing protons and  $\kappa \sim 4$ describing electrons (\cite{maks1997}) 

\begin{equation}\label{E:Eq7}
f(v) =\frac{n_{0}}{(\sqrt{\pi} w_{\kappa})^{3}} \frac{\Gamma(\kappa+1)}{\kappa^{3/2} \Gamma(\kappa-1/2)} (1+\frac{v^{2}}{\kappa w_{\kappa}^{2}}) ^{-(\kappa+1)}
\end{equation}

\noindent
where $w_{\kappa}=\sqrt{{(2\kappa-3)k_{B}T}/(\kappa m_{i})}$ is the most probable thermal velocity of a particle of mass $m_{i}$ and $\Gamma$ is the Gamma function. In a  $\kappa$-distribution the high velocity tail forms a larger fraction of the total ion population than in the M-B distribution and this distribution has a power-law decay (see \ref{E:Eq7}) instead of an exponential decay (see \ref{E:Eq5}). For the $\kappa$ distribution the escape flux fraction $F(w_{e})$ of particles with velocities greater than the escape velocity $w_{e}$ is given by (\cite{Pier2010})
   
\begin{equation}\label{E:Eq8}
F(w_{e}) = n_{0} A_{\kappa} w_{\kappa} (1+w_{e}^{2}/w_{\kappa}^{2})   (1+w_{e}^{2}/(\kappa w_{\kappa}^{2})) ^{-\kappa} \end{equation}

\noindent
where $A_{\kappa}\equiv(\Gamma(\kappa+1)/(4(\kappa-1)\kappa^{1/2}\Gamma(\kappa-1/2)\Gamma(3/2)))$. Note that the M-B distribution is derived from physically-grounded equilibrium statistical mechanics assumptions, while the $\kappa$ distribution is not, and instead is inferred from a fit to a somewhat poorly determined velocity distribution (see \cite{maks1997} and \cite{meye2007}, 5.4.2 and 5.5.2). However, the inference that the $\kappa$ distribution is therefore worse may be misleading because the assumption of thermal equilibrium may also be a poor one (\cite{meye2007} 5.4.2 and 5.7.6; this will be further discussed at the end of this section). 

\begin{figure}
         \includegraphics[width=\columnwidth]{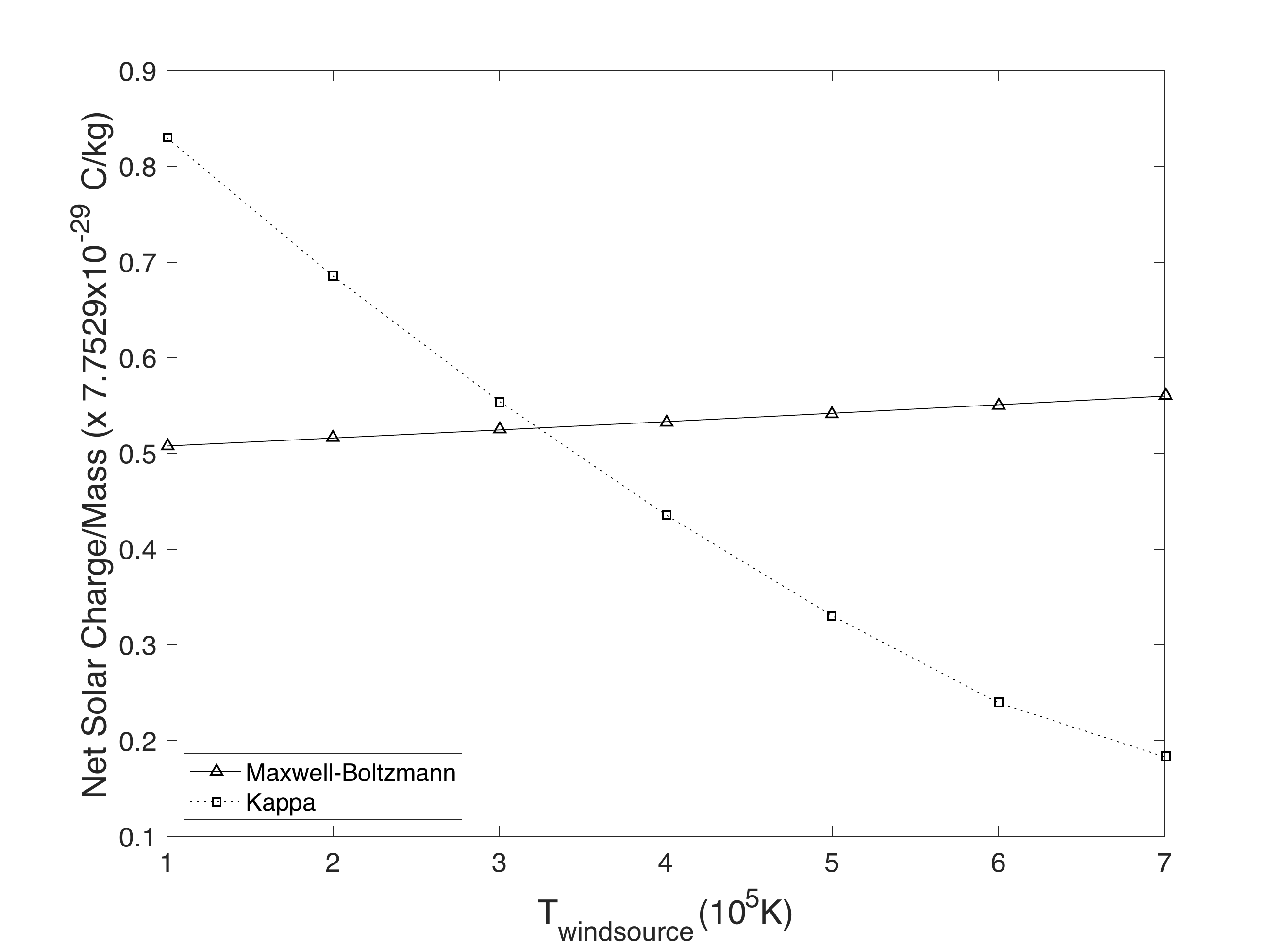}
         \caption{
         Net solar charge/mass ratio as a function of the temperature of the solar windsource that leads to equal numbers of thermal electrons and protons escaping in the solar wind from the corona of a trace-charged Sun.
Solid-line with triangle symbols shows escape in a solar wind composed of the Jeans tails of thermal Maxwell-Boltzman distributions for electrons and protons. Dashed-line with square symbols shows escape of a solar wind composed of the `Jeans-like' tail for the suggested parameterizations (Maksimovic et al., 1997)
of a thermal $\kappa$-distribution for electrons with $\kappa=$4, and the Jeans-tail of a thermal quasi-Maxwell-Boltzman distribution for protons,  approximated as the distribution $\kappa=$500. (see text).
}
   \label{fig:Fig2} 

\end{figure}

We can combine equations (2,3,6,8) to determine the solar charge/mass ratio needed for the solar wind to  have equal escaping fluxes of protons and electrons for a given windsource temperature. Figure 2 shows the corresponding predicted proton escape velocities for a thermal M-B and $\kappa$-distribution windsource near the base of the corona with `conventional' (\cite{meye2007}, Table 5.1)) windsource temperatures ranging from $1-\num{7e5}\si{K}$. For windsource temperatures $\sim \num{3.3xe5}\si{K}$ both the proposed M-B and $\kappa$ thermal windsource distributions predict a similar net solar charge/mass ratio $\codot \tcmuodotplus \simeq 0.53\tcmuodotplus$, and similar proton escape velocities of $\sim420 \si{km/s}$, e.g. $\sim 2/3$ the gravitational escape velocity of the Sun. The $\kappa$ distribution windsource predicts unreasonably low solar windspeeds for windsource temperatures much lower than $\num{2e5}\si{K}$, while the M-B distribution predicts only a small temperature and charge/mass dependence on the solar wind velocity. 

The trace-charged Sun hypothesis for the solar wind does not contradict Parker's kinematic model for the radial structure of the solar wind -- the only required change to Parker's original model is to replace the escape velocity for proton escape from the gravitational attraction of the sun with the escape velocity for proton escape from the electrogravitational attraction of the sun (Eqn. 2). For example, the stationary equilibrium state of a dynamic isothermal solar wind would still be described by combining equations (5+6) of \cite{park1965}:

\begin{equation}\label{E:Eq9}
v \frac{dv} {dr} \left(1- \frac{w_{i}^{2}} {v^{2}} \right) = -r^{2} \frac{d} {dr} \left( \frac{w_{i}^{2}} {r^{2}} \right) -\frac{w_{e}^{2}\Rsun} {r^{2}} \end{equation}

\noindent

Since the solar windsource temperature is a free parameter in Parker's model, the reduced net electrogravitational attraction of the sun on solar wind protons would lead to similar predicted windspeeds for $\sim50\%$ smaller  windsource temperatures.

\begin{figure}
	\includegraphics[width=\columnwidth]{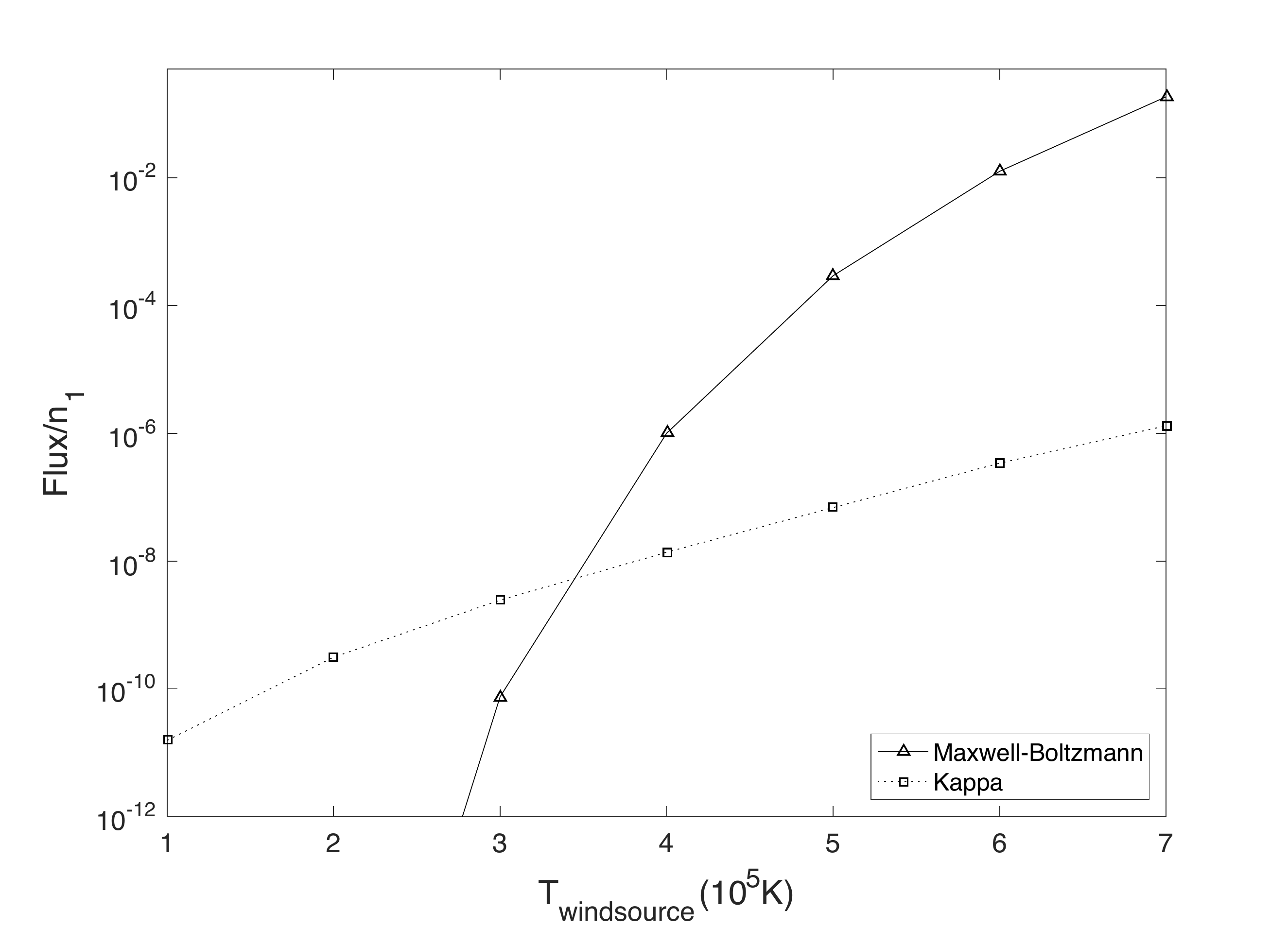}
         \caption{
Fraction of solar windsource protons with peak number density $n_{1}$  that is incorporated into the solar wind escaping from the base of the corona of a trace-charged Sun. (Note this wind has equal numbers of escaping protons and electrons, e.g. it is charge-neutral.)
Solid-line with triangle symbols shows flux fraction for a solar wind composed of the Jeans tails of thermal Maxwell-Boltzman distributions for electrons and protons. Dashed-line with square symbols shows escape of a solar wind composed of the `Jeans-like' tail for the suggested parameterizations (Maksimovic et al., 1997) of a thermal $\kappa$-distribution for electrons with $\kappa=$4, and the Jeans-tail of a thermal quasi-Maxwell-Boltzman distribution for protons, approximated as the distribution $\kappa=$500. Note that for a solar windsource temperature of 3 $\times$ 10$^{5}$, the two different particle distributions predict similar Jeans-like wind fluxes.  As discussed in the text, these fluxes will be augmented by $\sim 15 \times$ by the extra electron entrainment from escaping protons.
}
    \label{fig:Fig3SolarWindFluxes} 

\end{figure}

Figure 3 shows the predicted particle fraction of the windsource that is excited to escape velocity. For a M-B windsource velocity distribution, the predicted solar windflux is a strong function of windsource temperature. It is  a much weaker function of windsource temperature for the proposed proton/electron $\kappa$ distributions (Figure 3). Again the M-B and $\kappa$ statistics scenarios predict similar fluxes of $10^{-9}$ times the windsource particle number density $n_{1}$ for windsource temperatures $\sim \num{3.3e5}\si{K}$. However, below we will see that these estimated fluxes are likely to be at least an order of magnitude too small.

In this scenario, the solar wind has equal numbers of escaping electrons and protons because the sun retains a minute net residual positive charge that remains from its original net positive charge after accretion and starbirth. The net positive charge induces an electric field that attracts electrons and repels protons, so that protons feel a net electrogravitational attraction to the sun about half their gravitational attraction, while electrons feel a net electrogravitational attraction about 1000 times their gravitational attraction. This is why equal numbers of electrons and protons escape from the Sun's corona even though lower-mass electrons typically have much higher speeds (e.g., Eq. 5) due to their $\sim 43$ times faster mean thermal velocity than that of $\sim$1836 times more massive protons. To explain equal numbers of electrons and protons in the solar wind, solar wind researchers have proposed that there is an external electric field that preferentially `pulls' protons from the sun (\cite{jock1970,holl1970,lema1971}), refining an earlier idea of \cite{Pann1922} and \cite{Ross1924}. The idea is that a surplus of electrons would tend to initially escape with the solar wind due to lower mass electrons having a much larger high-velocity tail with speeds greater than the Sun's gravitational escape velocity. This preferential electron escape, in turn, would induce a net negative charge imbalance near the corona, resulting in an additional local electric field that preferentially pull protons from the Sun -- up until the solar wind becomes essentially charge-neutral (\cite{jock1970,holl1970,lema1971}). 

I think the gist of this argument is correct. My revised perspective is that, for a minutely-charged sun, the naturally escaping Jeans-tail of protons is likely to entrain a similar flux of negligible-mass electrons with entrained speeds much less than the escape velocity for a free thermal electron. Each escaping proton that entrains an electron is not available to balance the electric field induced by escaping free electrons. We know that escaping protons somehow entrain a significant fraction of four times more massive $\ce{He}$ nuclei, with the solar wind's $\ce{He}/\ce{H}$ ratio being roughly half the sun's bulk $\ce{He}/\ce{H}$, even though the predicted thermal Jeans-tail for escaping $\ce{^{4}He++}$ is minuscule in comparison to the flux of escaping $\ce{^{1}H+}$.  If escaping $\ce{^{1}H+}$ entrain $0.9e^{-}$ per $\ce{^{1}H+}$ ion, the induced Lemaire-Scherer-Junkers-Hollweg (LSJH) electric field would lead to a 10-fold increase of the solar windflux above the Jeans-flux estimate shown in Figure 1c, if $0.99$, a 100-fold increase, etc. The consequence for electron velocity distributions is that the entrained electrons should share the proton velocity distribution (e.g. have a velocity distribution like the Maxwellian tail of the escaping protons), which would resemble a much cooler thermal electron distribution for a pseudo-temperature $T_{e^{-}} = T_{\ce{H+}}(m_{e}/m_{\ce{H+}})$. To this `bulk velocity' of the escaping electrons would be added the Maxwellian of the actual higher-temperature electron Jeans-escape tail. The ratio of high-energy-tail `halo' electrons to main energy `entrained-by-escaping-protons' electrons would reflect the ability of escaping protons to entrain electrons as they escape. This provides an attractive physical picture to interpret the observed `double-Maxwellian-like' electron velocity distribution within the solar wind. (In contrast, escaping protons are not observed to have a high-energy `double-Maxwellian' or $\kappa$ `halo' to their Maxwellian-like velocity distribution.) The observed low-velocity fraction of electrons has a pseudo-thermal temperature of $\num{1.6e5}K$ and $n_{0}=30.8/cm^{3}$, while the high-speed halo is fit by the tail of a M-B thermal distribution with a temperature of $\num{8.9e5}K$ and $n_{0}=2.2/cm^{3}$ (\cite{maks1997}). These observations would suggest that the LSJH fraction of escaping electrons and protons is $30.8/2.2 = 15.4$ times the electron and proton fraction that would naturally escape by the Jeans thermal escape mechanism (e.g. 15.4 less-energetic proton-entrained electrons escape for each naturally escaping free thermal electron). However, while attractive in concept, the implied double-Maxwellian temperatures do not differ by the factor implied by the simplest form of this argument. While this mechanism may be responsible for changing the escaping electron and proton velocity distributions,  a proper quantitative fit appears to require a more sophisticated theoretical treatment than the extreme simplifications that were used in this exploratory analysis.

Among its major simplifications, the above exploration assumed an idealized spherical solar wind and did not include magnetic effects. Note that the magnetic field component due to the rotation of the nearly uncharged Sun is minute, on the order of $\SI{5e-22}{T} \; ($\SI{5e-18}{Gs} ) at the Sun's surface where the Sun's background polar magnetic field is of order 1 Gs. This estimate uses equation (13) below developed to assess the magnetic field of a trace-charged neutron star, with the Sun's application assuming its effective mass lies within its radiative zone ($0.7 {\Rsun}$ containing 97\% of the Sun's mass), that its radiative zone is rotating at $432nHz$, and that the Sun's charge-to-mass ratio is  $\simeq 0.53\tcmuodotplus$.

Because the detailed physics for the generation of the solar wind is thought to involve complex, poorly quantified and poorly understood magnetohydrodynamic processes (\cite{meye2007}, section 5.6.2) within a Sun with a strong magnetic field, and because solar wind researchers already invoke electric fields arising from the LSJH effect to interpret observed characteristics of the solar wind, I anticipate that the trace-charged Sun hypothesis will likely be controversial within this field.  In contrast, the current enigmas of the strong (98\%) observed predominance of protons and positive ions in cosmic rays (\cite{bere1990,bhat2000}), and the extraordinary energies involved in observed axial jetting away from active galactic nuclei (cf. Figure 4) (\cite{ghis2014}) and pulsars do not have currently well-accepted theories to explain them. Here the trace-charged matter hypothesis should have a lower threshold to satisfy in order to be recognized as a potential improvement to currently favored ideas.

\begin{figure}
	\includegraphics[width=\columnwidth]{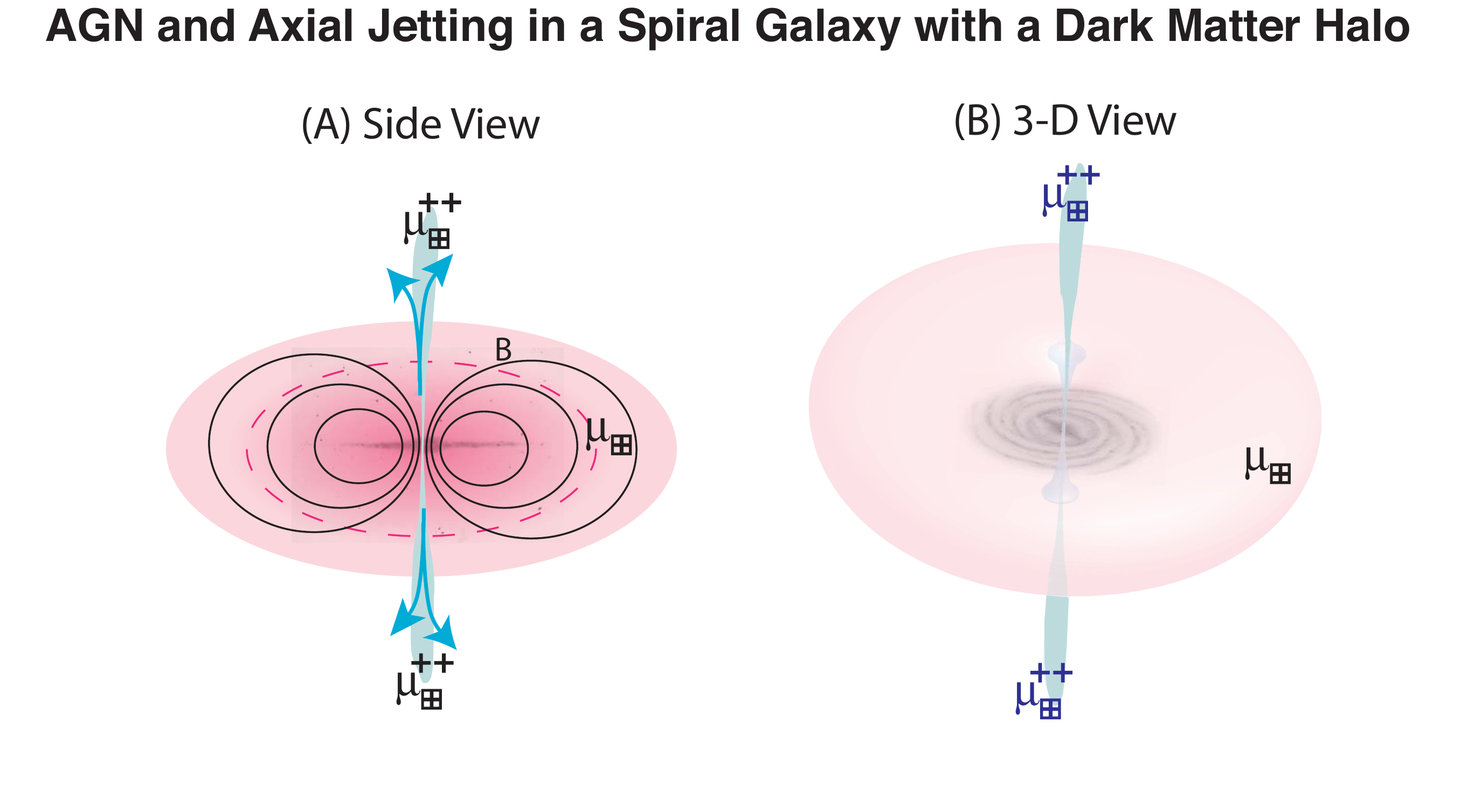}
         \caption{
Sketch illustrating (A) side-view cross-section and (B) 3-D perspective view of the scenario for the generation of cosmic rays by expulsion of protons from near the event horizon of a trace-charged black hole, here an ultramassive trace-charged black hole proposed to sit at the center of an `Active Galactic Nucleus' (AGN) This ultramassive trace-charged black hole is shown embedded within a spiral galaxy. It leads to jetting dark matter $\tcmuboxplusplusplus$ (shown by blue clouds) that are a source of cosmic rays (see text). Axial Jetting occurs along magnetic dipole field-lines of the trace-charged black hole that are shown by black lines annotated with `B'. Also drawn is the dark-matter $\tcmuboxplus$ halo (pink cloud) to this galaxy that is composed of unseen clouds of isothermal trace-charged dark matter in the form of [$\ce{H2}+$ trace $\ce{H3+}$] molecular clouds. Luminous matter within the galaxy is shown by the central grey outline. The dark-matter halo's presence is detected by the flat rotation curve of the distal radiovisible $\sim$12\% local mass fraction $\tcmuboxdot$-HI clouds  that are embedded within an unseen matrix of isothermal trace-charged dark matter [$\ce{H2}+$ trace $\ce{H3+}$] molecular clouds.  (See Figure 5 and text.).
}
    \label{fig:Fig4AGNJettingHaloEPS} 

\end{figure}

\section{Trace-charged dark matter interpretation of axial jetting from AGN and Black Holes}

Unlike warm or bright matter, a black hole will keep its charge as it accretes by swallowing trace-charged matter. The resulting $\epsCG$-like charge/mass ratio of black holes, including the ultramassive black holes that power active galactic nuclei (AGNs), can lead to the formation of strong Coulomb-force-driven axial jets if atoms ionize as they approach the black hole's event horizon. Synchrotron/bremsstrahlung radiation from AGN jets would be energetic enough to be the source powering the continuum radiation of quasars and blazars.

Imagine the case of a positively trace-charged black hole that rotates, thereby generating a dipolar magnetic field. If accreting atoms ionize at a distance $aR_{S}$, where $R_{S}$ is the Schwarzschild radius of a black hole, free electrons will feel a strong Coulomb attraction to a black hole with charge/mass ratio $\cBH \epsCG$, while protons will be repelled. The electrostatic work done during the proton's repulsion along magnetic field lines can easily accelerate protons and their entrained matter to the relativistic speeds observed  in axial jetting from AGNs.

The electrostatic work $U_{\scriptscriptstyle{BH}}^{p^{+}}(r)$ done on jetting protons and their entrained matter is

\begin{equation}\label{E:Eq10}
U_{\scriptscriptstyle{BH}}^{p^{+} }= \frac {-k_{E}Q|e^{-}|} {r} \biggm |_{aR_{S}}^{\infty} = \frac{+k_{E}(\cBH \epsCG M)|e^{-}|}{aR_{S}}
\end{equation}

\noindent
where the black hole has mass $M$ and charge $Q\sim \cBH \epsCG M$ ($|\cBH| \le 1$).
The work from the black hole's electrostatic repulsion of escaping protons can be further simplified using the relation $R_{S}=2MG/c^{2}$, where $M$ is the mass of the black hole and the speed of light $c=\SI{3e8}{m/s}$.

\begin{align}\label{E:Eq11}
U_{\scriptscriptstyle{BH}}^{p^{+}} &= \frac {k_{E}\cBH \epsCG|e^{-}|c^{2}} {2Ga}\\
&= \num{8.35e7}\frac{\cBH}{a} \left[\frac{J}{p^{+}} \right] = \num{5.21e26}\frac{\cBH}{a} \left[ \frac{eV}{p^{+}} \right] \notag
\end{align}

\noindent
Axial jetting from the M87 AGN has been observed to start from at least $5.5R_{S}($\cite{Doel2012}), e.g. $a\le 5.5$. For $\cBH/a \sim 0.2$, this would imply that jetting protons would have $\sim \SI{1e26}{eV/p^{+}}$ of electrostatic work done on them as they are ejected from $\sim5.5R_{S}$. This work is transformed into kinetic energy of the jetting material.  Their kinetic energy depends on their relativistic speed measure (Lorentz factor) $\gamma = 1/\sqrt{1-v^{2}/c^{2}}$ as

\begin{equation}\label{E:Eq12}
(\gamma-1)m_{p}c^{2} = \num{5.21e26}\frac{\cBH}{a} \left[ \frac{eV}{p^{+}} \right]
\end{equation}

\noindent
This energy, if supplied to accelerate a single proton to its maximum possible speed is enormous, of order $\sim1\%$ of the Planck energy  $\SI{1.22e28}{eV}$, and of order $\num{1e6}$ times the GZZ energy cutoff  $\SI{3.12e20}{eV}$ (\cite{grei1966,zats1966}) for cosmic rays that interact with a significant number of CMB photons en route to Earth. It would imply that any axial jetting from a trace-charged black hole in the Galaxy or a nearby galaxy could create cosmic ray protons that arrive with energies in excess of the GZZ limit, as is possibly observed(\cite{abre2010}).

It seems likely that this energy could be used to collisionally accelerate a larger number of bulk neutral charge electrons and ions to smaller, yet still relativistic speeds up to $\sim 0.99c$ as observed in some AGN jets. A particle moving at $0.99c$ has $\gamma\approx7$. The kinetic energy of a proton moving at $0.99c$ is roughly $6m_{p}c^{2}$ or $\sim6GeV$. This implies that a single proton ejected from a distance $5R_{S}$ could collisionally accelerate $\num{1e17}$ additional hydrogen atoms to $0.99c$. Restated, ejecting the `extra' protons from $\sim10\%$ of infalling dark matter with a net charge ratio of $1$ excess proton per $\num{1.12e18}$ hydrogen atoms could lead to axial jets containing $\sim1\%$ of infalling material that are ejected at $0.99c$ with a trace-charge ratio of $1$ excess proton per $\num{1e17}$ hydrogen-atom equivalent ions, $10 \times \epsCG$. 

This scenario (see Figure 4) appears to be a straightforward physical mechanism for the creation of strong axial jets when matter ionizes as it accretes onto a rotating trace-charged black hole. It is also a potential way by which a trace-charged black hole can keep its net charge-to-mass ratio below $\epsCG$ so that it can continue to attract and consume surrounding clouds of trace-charged dark matter, while, at the same time, increasing the net charge-to-mass ratio in its surrounding regions.

This AGN mechanism also offers a qualitative explanation of the currently enigmatic observation that star birth in galaxies slows once an AGN forms (\cite{page2012}). Once formed, the AGN could 'spike' surrounding regions with ejected $\tcmuboxplus$, thereby increasing their resistance to electrogravitational collapse and star formation.  It also appears that this mechanism can create relativistic jets with larger luminosity that the black hole's accretion disk (\cite{ghis2014}). There is no need for the axial jets around an AGN to have the same speeds -- a paradox observation for some theories of jet formation (\cite{meye2015}) -- since individual jet speeds would depend upon the net rate and charge/mass ratio of material accreting onto a given hemisphere of the black hole as well as the net charge/mass ratio $\cBH \epsCG$ of the black hole.

\section{Neutron stars, pulsars, and magnetars}

Unlike less dense forms of warm or bright matter, neutron stars, pulsars, and magnetars should keep their initial and accreted charge, because their strong force attractions between protons and neutrons are much stronger than the repulsive Coulomb forces created by $\epsCG$-like excesses of trace positive charge. Conventionally, these stars are thought to form by stellar core collapse during a supernova (\cite{lyne2012}). Since a supernova explosion event will excite electrons to $\sim 43 \times$ the mean Maxwell-Boltzmann thermal speed of protons and neutrons, it seems plausible that an electron-rich outer shell would be preferentially ejected during the supernova explosion, consistent with the observed distribution of trace electron synchrotron radiation in the outer regions of the still-expanding Crab Nebula remnants (e.g. \cite{bran2008}). Residual supernova core material would have a corresponding initial trace surplus of protons and ions to electrons, leading to the neutron star's initial material having a slight excess of protons to electrons within the $\sim 5\%$ of electrons/protons that reside within the star's ultradense, neutron-dominated, nucleon soup. (This is simply the conventional hypothesis for the birth of a neutron star/pulsar/magnetar (\cite{lyne2012}), augmented by the hypothesis that a trace charge imbalance is also created during its supernova-linked genesis).

A trace-charged neutron star with a charge-to-mass ratio $\epsCG$ should have axial jetting similar to that from near the event horizon of a black hole. Such axial jetting has been documented in the X-ray spectrum for the Crab Nebula and Vela pulsars (\cite{weis2000,pavl2003,lyne2012}). 

The background dipole field of a rotating trace-charged neutron star can be approximated as that of a uniform density,uniform charge density rotating sphere. Assume the sphere has a radius $R$, angular rotation rate $\omega$, mass $M$, and net charge $Q= \tcmuCG M$. 
The dipole field $B(r,\theta)$ at a distance $r$ and colatitude $\theta$ from the center of the spin axis of this sphere of charge is

\begin{equation}\label{E:Eq13}
B(r,\theta) = \frac{\mu_{0} \omega M \tcmuCG R^{2}} {20 \pi r^{3}}  \big( 2 cos(\theta) \hat{r} + sin(\theta) \hat{\theta} \big)
\end{equation}

A typical pulsar is thought to have a radius $R \sim \SI{10}{km}$, mass $M=1.4 \Msun$, and rotation rate $\omega \sim \SI{1}{s^{-1}}$. For these parameters, the estimated polar magnetic field strength is $\SI{9.6e8}{T}$  ($\SI{9.6e12}{Gs}$), in the upper range of typical  $\num{1e7}-\SI{1e9}{T}$  ($\num{1e11}-\SI{1e13}{Gs}$) pulsar field-strengths estimated from the observed absorption lines (\cite{bign2003}), pulsed radio emissions (\cite{cami2006}), and spin-down rates of pulsars (\cite{olau2014}), and $1-2$ orders of magnitude less than the maximum magnetic field strength estimates for magnetars (\cite{olau2014}).  Magnetars would need to have either greater mass and charge than typical pulsars and/or a stronger additional dynamo component to their field. A typical pulsar would also need to have an additional dynamo or frozen field component to explain the offset between the pulsar's spin and magnetic poles.

The hypothesis that pulsars contain a trace positive charge can also explain their inferred increase in magnetic field strength as they age (\cite{lyne2012}). If trace positive charge is initially well-mixed throughout the nucleon soup, and Coulomb repulsion leads to net outward migration of trace positive charge (or net inward migration of charge-balancing electrons), then its spin-induced magnetic field strength would increase. The accelerating polar jet of protons, ions, and entrained electrons would also generate broad spectrum synchrotron/bremsstrahlung radiation. Could this be the source of X-ray emissions from the magnetic polar regions of pulsars?

\section{Trace-charged dark matter Galaxy Halos}

The most unequivocal evidence for the existence of intragalactic dark-matter is from observations of the circular velocities of HI-clouds (HI $\equiv \ce{H1}$) in spiral galaxies as a function of radius (\cite{fabe1979,Bahc2014,Bahc2015}) (see Figure 5). Radiotelescopes can observe the bulk distribution and velocity of neutral  HI clouds in many spiral galaxies from their $21-cm$ radio emissions.  Frequently these clouds have a near-uniform rotation speed with increasing radius (\cite{rubi1970,bege1989,sofu2001}) that does not decay with $r^{1/2}$ as would be expected for the mass contained within the more central luminous matter within the galaxy. Instead the rotation curve remains relatively flat to the maximum radius where HI-clouds can be observed, suggesting there is unseen matter whose density decreases radially proportional to $r^{-2}$. In this case, the additional gravitational attraction from this dark-matter on HI-clouds would lead to their near-uniform rotation speeds as a function of radius.

\begin{figure}
	\includegraphics[width=\columnwidth]{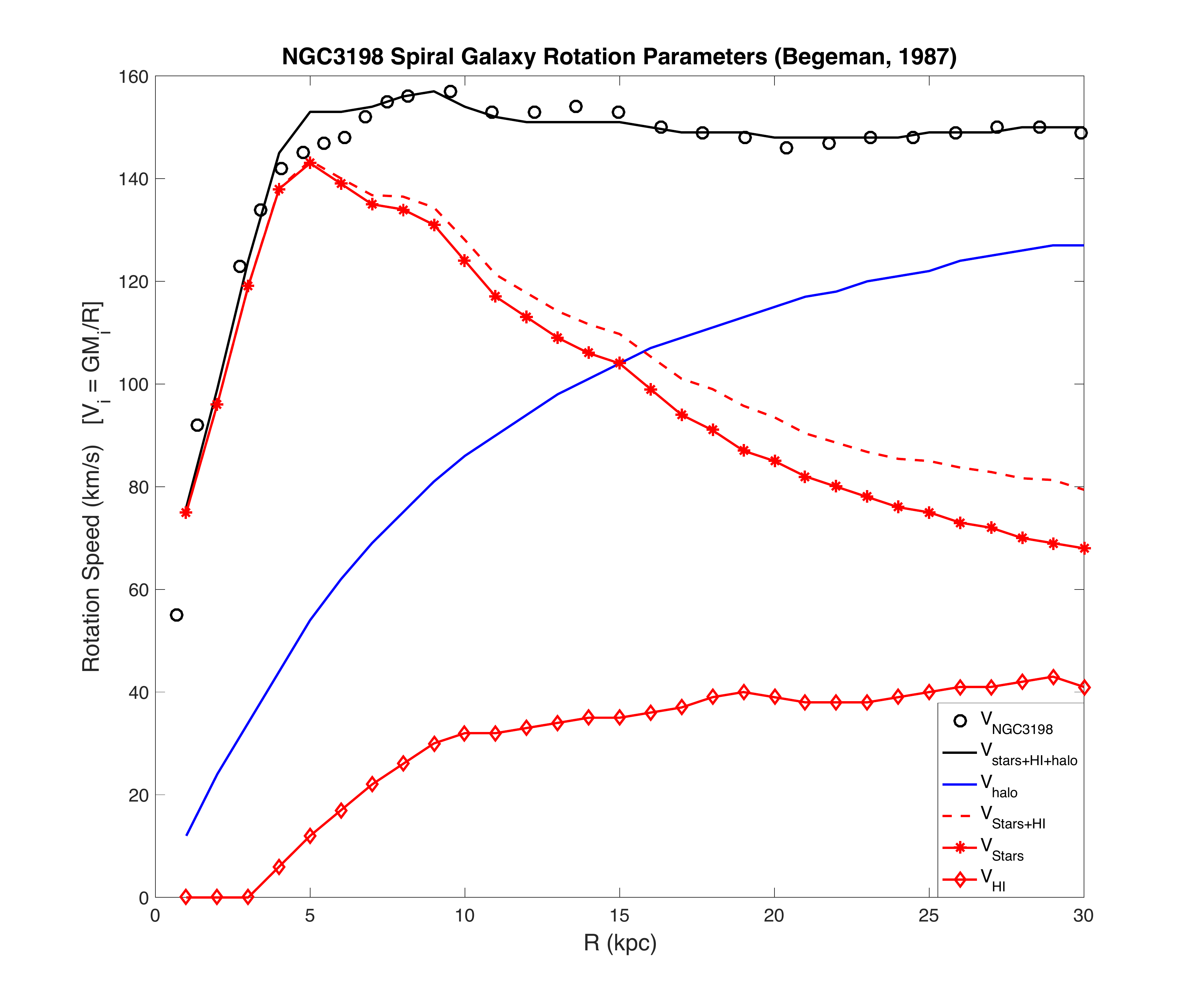}
         \caption{
         Observed rotation curve $vs.$ radial distance R (black circles, observations $\pm$3 kms$^{-1}$) for the spiral galaxy NGC3198 (Begeman et al., 1989) compared to predicted rotation curves for luminous stellar material (red stars), radioluminous $\tcmuboxdot$-HI gas clouds (red diamonds), and Begeman's (1987) preferred radial density distribution for a Bonnor-Ebert isothermal dark matter halo that is proposed here to be composed  of trace-charged $\tcmuboxplus$- $\ce{H2}$ gas clouds (solid blue line). The rotation curve predicted for all luminous and radioluminous mass is shown as the red dotted line. Clearly a large component of unseen halo matter is required to fit the observed rotation curve -- this is the well-known 'Missing Dark Matter' conundrum of the apparent non-Keplerian rotation of the distal $\tcmuboxdot$-HI gas clouds within spiral galaxies.
         }
    \label{fig:Fig5NGC3198Rotation} 

\end{figure}

I propose this dark matter is  $\tcmuboxplus$ in the form of clouds of trace-charged molecular hydrogen. Small $\tcmuboxdot$-HI clouds with a volume fraction $\fboxdot \lesssim 0.25$  embedded within a much larger volume fraction of [$\ce{H2}+$ trace $\ce{H3+}$] molecular clouds can generate an isothermal-sphere-like radial mass distribution of $\tcmuboxplus$-$\ce{H2}$ that will lead to near-uniform $\tcmuboxdot$-HI cloud rotations. Next will be a brief outline of the equations governing an isothermal-sphere-like $\tcmuboxdot$-HI + $\tcmuboxplus$-$\ce{H2}$ radial mass distribution. Then I will compare observations from the type-example NGC3198 spiral galaxy against predictions of this theory, and show that its observed rotation curve is consistent with the matter in its halo being composed of a two-phase isothermal sphere consisting of $\sim$12\% by mass ($\sim$23\% by volume) $\tcmuboxdot$-HI clouds  embedded within $\sim$88\% by mass ($\sim$77\% by volume) [$\ce{H2}+$ trace $\ce{H3+}$] molecular clouds, both with cloud densities decaying radially as $r^{-2}$. 

The theory assumes that neighboring clouds of $\tcmuboxdot$-HI and $\tcmuboxplus$- $\ce{H2}$ are in local pressure and temperature equilibrium.  The basic equations describe hydrostatic equilibrium within a mixture of $\tcmuboxdot$-HI and $\tcmuboxplus$-$\ce{H2}$ gas clouds. The two types of gas clouds form discrete subregions that are in local pressure-temperature equilibrium with each other. Each subregion is described by the ideal gas equation, e.g. 

\begin{equation}\label{E:Eq14}
P = \frac{\rhoplus k_{B}T}  {\moverbarplus m_{p}} = \frac{\rhodot k_{B}T}  {\moverbardot m_{p}} 
\end{equation}

\noindent
where $\moverbardot m_{p}$ and $\moverbarplus m_{p}$ are the mean particle masses of $\tcmuboxdot$ and $\tcmuboxplus$, respectively, with the proton mass $m_{p}$ being a convenient reference unit.  In this case $\rhoplus / \moverbarplus = \rhodot / \moverbardot$ or $  \rhodot = \rhoplus (\moverbardot / \moverbarplus) \simeq 0.53 \rhoplus$ for intragalactic matter of bulk composition $\ce{H_{1}He_{0.075}O_{5.5e-4}C_{3.5e-4}N_{8.5e-5}}$.   Assume we have measurements of the cumulative mass fraction $\ML (R)$ due to mass contained within the stars and central bulge of the spiral galaxy. Then the equations describing the bulk hydrostatic equilibrium of this mixture are

\begin{align}\label{E:Eq15}
 \frac{dP(R)}{dR} = &\frac{1}{R^{2}}  \lbrace  ( k_{E}(\epsCG)^{2} \cGBH \cboxplus - G )  \MGBH \\
                            &- G( \ML (R)+ \Mboxdot (R)) \notag\\
                            &+ (  (k_{E} ( \epsCG )^{2} (\cboxplus)^{2}  -G) \Mboxplus (R) \rbrace (1-\fboxdot ) \rhoplus (R) \notag\\ 
                            &- \frac {G} {R^{2}} \lbrace \MGBH + \ML (R) + \Mboxdot (R) + \Mboxplus (R) \rbrace \fboxdot \rhodot (R) \notag
\end{align}

\noindent
where $\MGBH$ and $\cGBH$ are the mass and trace-charge fraction of a possible central galactic black hole, and the cumulative mass distributions $\Mboxplus (R)$ and $\Mboxdot (R)$ are defined by

\begin{align}\label{E:Eq16}
 \frac {d \Mboxplus (R)} {dR} &= 4 \pi R^{2} \left( 1-\fboxdot \right) \rhoplus (R)\\
 \frac {d \Mboxdot (R)} {dR}  &= 4 \pi R^{2} \fboxdot \rhodot (R) = 4 \pi R^{2} \fboxdot  \left( \frac{\moverbardot}{\moverbarplus} \right) \rhoplus (R)
 \end{align} 

\noindent
The cumulative charge $\Qboxplus(R)$ associated with dark matter clouds is

\begin{equation}\label{E:Eq18}
\Qboxplus(R) = \cboxplus \epsCG \Mboxplus(R)
\end{equation}

\noindent
To simplify equation (15), I will assume that the $\tcmuboxdot$-HI gas fraction $\fboxdot$ is uniform for the region of interest ($R>R_{L}$), and will also substitute $\rhodot = \rhoplus (\moverbardot / \moverbarplus)$.  Then, for this region

\begin{align}\label{E:Eq19}
  \frac{dP(R)}{dR} = &\frac{(1-\fboxdot)G}{R^{2}}  \Big \lbrace  (\cGBH \cboxplus - 1 )  \MGBH - ( \ML (R)\\
                            &+ \Mboxdot (R)) +  ((\cboxplus)^{2}  -1) \Mboxplus (R) \Big \rbrace  \rhoplus (R) \notag\\ 
                            &- \frac {\fboxdot G} {R^{2}} \Big \lbrace \MGBH + \ML (R) + \Mboxdot (R) + \Mboxplus (R) \Big \rbrace  \frac {\moverbardot} {\moverbarplus} \rhoplus (R) \notag
 \end{align} 

\noindent
To further simplify for initial analysis, I will only examine the case where $\cboxplus \simeq \cGBH \simeq 1$. This leads to

\begin{align}\label{E:Eq20}
  \frac{dP(R)}{dR} = &-\frac{G}{R^{2}}  \Bigg \lbrace  \frac {\fboxdot \moverbardot } {\moverbarplus } \MGBH\\ 
  &+  \left( \frac {\fboxdot \moverbardot } {\moverbarplus } 
  + \left( 1-\fboxdot \right)  \right) \left( \ML (R) + \Mboxdot (R) \right) \notag\\
  &+ \frac {\fboxdot \moverbardot } {\moverbarplus }  \Mboxplus (R) \Bigg \rbrace  \rhoplus (R) \notag
 \end{align} 

\noindent
which can be rewritten as 

\begin{align} \label{E:Eq21}
  \frac {dP(R)}{dR} = & - \frac {G}{R^{2}}  \Bigg \lbrace  \fboxdot  \frac {\moverbardot } {\moverbarplus } \MGBH \\
  &+  \left( 1- \fboxdot  \left( 1- \frac {\moverbardot } {\moverbarplus } \right)   \right)  \left( \ML (R) + \Mboxdot (R) \right) \notag\\
  &+  \fboxdot  \frac {\moverbardot } {\moverbarplus } \Mboxplus (R)  \Bigg \rbrace  \rhoplus (R) \notag
  \end{align} 

\noindent
Next we assume the clouds behave as isothermal ideal-gases, with their pressure-density relation relation expressed in terms of the isothermal speed of sound $\csplus$ within the $\tcmuboxplus$-$\ce{H2}$ gas:

\begin{equation} \label{E:Eq22}
\csplus  = \sqrt{ \frac{\rhoplus k_{B}T}  {\moverbarplus m_{p}} }
\end{equation}

\noindent
This relation is used to transform the radial pressure derivative into a radial derivative of density:

\begin{equation} \label{E:Eq23}
\frac{dP(R)}{dR} = {\csplussquared} \frac  {d\rhoplus(R)}{dR} 
\end{equation}

\noindent
For regions outside the maximum radius $R_{L}$ of the galaxy stars, e.g. $R>R_{L}$ , equation (21) becomes

\begin{align}\label{E:Eq24}
{\csplussquared} \frac  {d\rhoplus(R)}{dR} 
= & - \frac {G}{R^{2}}  \Bigg \lbrace  A(R_{L})
  +  \left( 1- \fboxdot  \left( 1- \frac {\moverbardot } {\moverbarplus } \right)   \right) 
 \Mboxdot (R)\\
 &+  \fboxdot  \frac {\moverbardot } {\moverbarplus } \Mboxplus (R)  \Bigg \rbrace  \rhoplus (R) \notag
  \end{align} 

\noindent
where 

\begin{align}\label{E:Eq25}
A(R_{L}) =  & \Bigg \lbrace  \fboxdot  \frac {\moverbardot } {\moverbarplus } \MGBH 
  +  \left( 1- \fboxdot  \left( 1- \frac {\moverbardot } {\moverbarplus } \right)   \right) \times \\
  &\left( \ML (R_{L}) + \Mboxdot (R_{L}) \right) +  \fboxdot  \frac {\moverbardot } {\moverbarplus } \Mboxplus (R_{L})  \Bigg \rbrace \notag
  \end{align} 

\noindent
and

\begin{align}\label{E:Eq26}
\Mboxdot (R) =  \Mboxdot (R_{L}) + \Bigg \lbrace  \frac {\fboxdot  \frac {\moverbardot } {\moverbarplus }}  {1- \fboxdot } \left(  \Mboxplus (R) - \Mboxplus (R_{L})  \right)   \Bigg \rbrace 
 \end{align} 

\noindent
For $\fboxdot<<1$ and $R>R_{L}$, equations (23+25) can be further simplified to

\begin{align}\label{E:Eq27}
{\csplussquared} \frac  {d\rhoplus(R)}{dR} 
= & - \frac {G}{R^{2}}  \Bigg \lbrace  A(R_{L}) + 2 \fboxdot \frac {\moverbardot } {\moverbarplus } \Mboxplus (R) \Bigg \rbrace  \rhoplus (R) 
 \end{align} 

\noindent
Equations (16) and (27) describe the hydrostatic equilibrium of a $\tcmuboxdot$-HI + $\tcmuboxplus$-$\ce{H2}$ clouds with a small fraction of  $\tcmuboxdot$-HI relative to `dark' trace-charged $\tcmuboxplus$-$\ce{H2}$ matter. These equations have the same mathematical form as the equations governing the hydrostatic equilibrium of an uncharged isothermal gas with a much weaker effective gravitational constant $G_{eff} \simeq 2 \fboxdot \frac {\moverbardot } {\moverbarplus }G$:

\begin{align}\label{E:Eq28}
\frac{dP(R)}{dR}    &=  - \frac {G_{eff} M(R)}{R^{2}}  \rho (R)\\
\frac {d M(R)} {dR} &= 4 \pi R^{2} \rho (R)
\end{align} 

\noindent
which means that the $\rho (R) \propto R^{-2}$ distal region of the analytical solution for a Bonnor-Ebert isothermal self-gravitating sphere will, with appropriate reparameterization for an effective weaker self-attraction, also describe the distal behavior of the gas-clouds in this system. Here I choose to use the analytical distribution assumed in Begemen's textbook example of the dark-matter halo for NGC1398.  This simple closed-form approximation to the distal NGC1398 dark-matter distribution is

\begin{equation} \label{E:Eq30}
\rhoplus (R) = \rho_{0} \left(1+ \left( \frac {R} {R_{C}} \right)^{2} \right)^{-1}
\end{equation}

\noindent
with the rotational velocity $V_{C}$ of the corresponding dark-matter-only rotation curve given by

\begin{equation} \label{E:Eq31}
V_{C}^{2} (R) = 4 \pi G \rho_{0} R_{C}^{2} \left[ 1 - \frac {R_{C}} {R} \arctan \left( \frac {R} {R_{C}} \right) \right]
\end{equation}

\noindent
Begeman's best-fit parameters for the density distribution of the dark-halo of NGC3198 are $R_{C} = \SI {7.5} {kpc} = \SI{2.3e20} {m}$,  $V_{max}^{2} (R) = (150$kms$^{-1})^{2}$, and $\rho_{0} = \SI {5e-22} {kgm^{-3}}$ (\cite{bege1987}). (Note that the gravitational attraction between $\tcmuboxplus$ matter and $\tcmuboxdot$-HI clouds is `unchanged' normal gravitation at strength $G$, thus this new analysis leads to the same approximate dark-matter halo distribution as that estimated by previous workers (\cite{bege1987}.)

\begin{figure}
	\includegraphics[width=9cm]{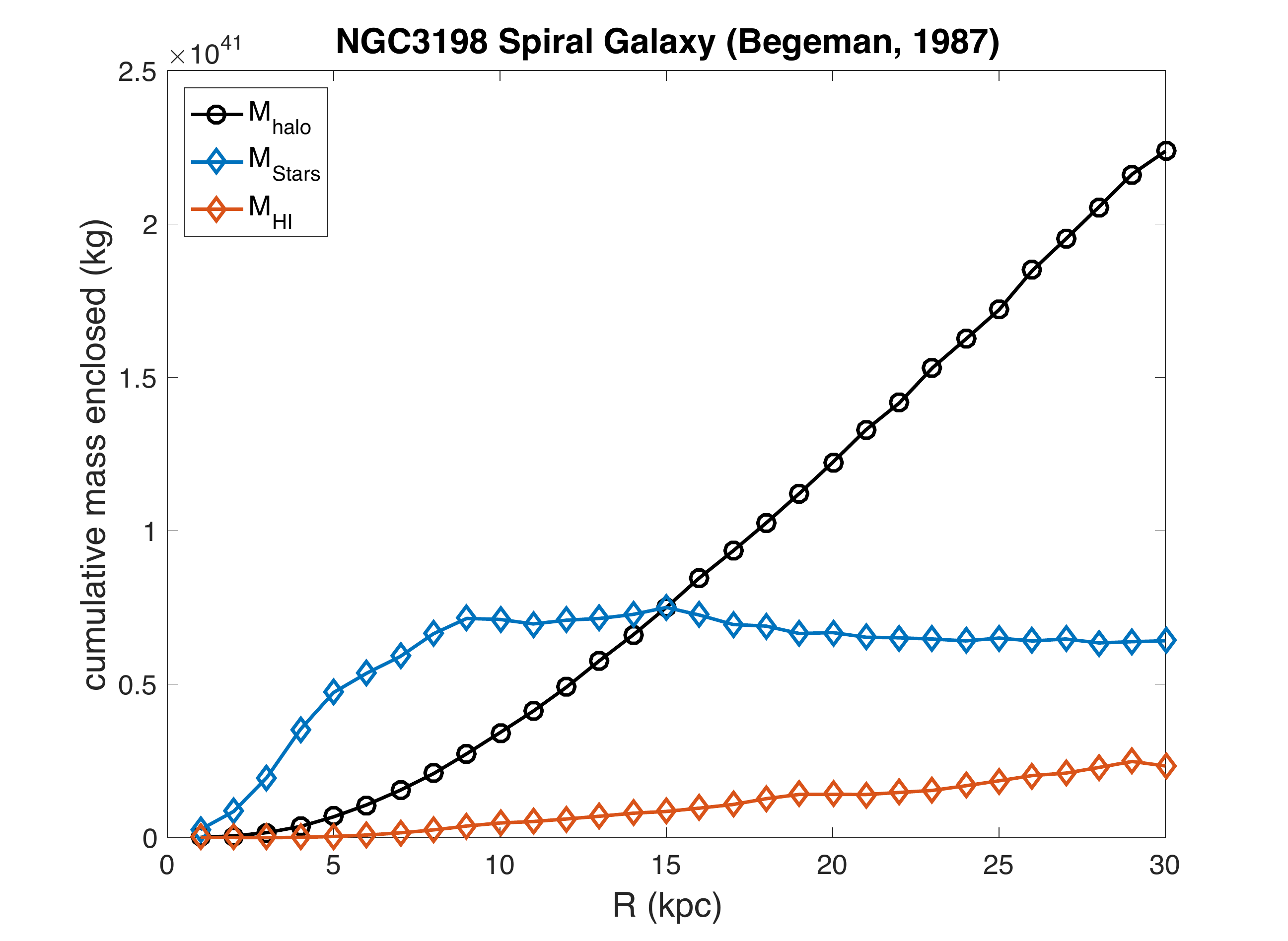}
        \includegraphics[width=9cm]{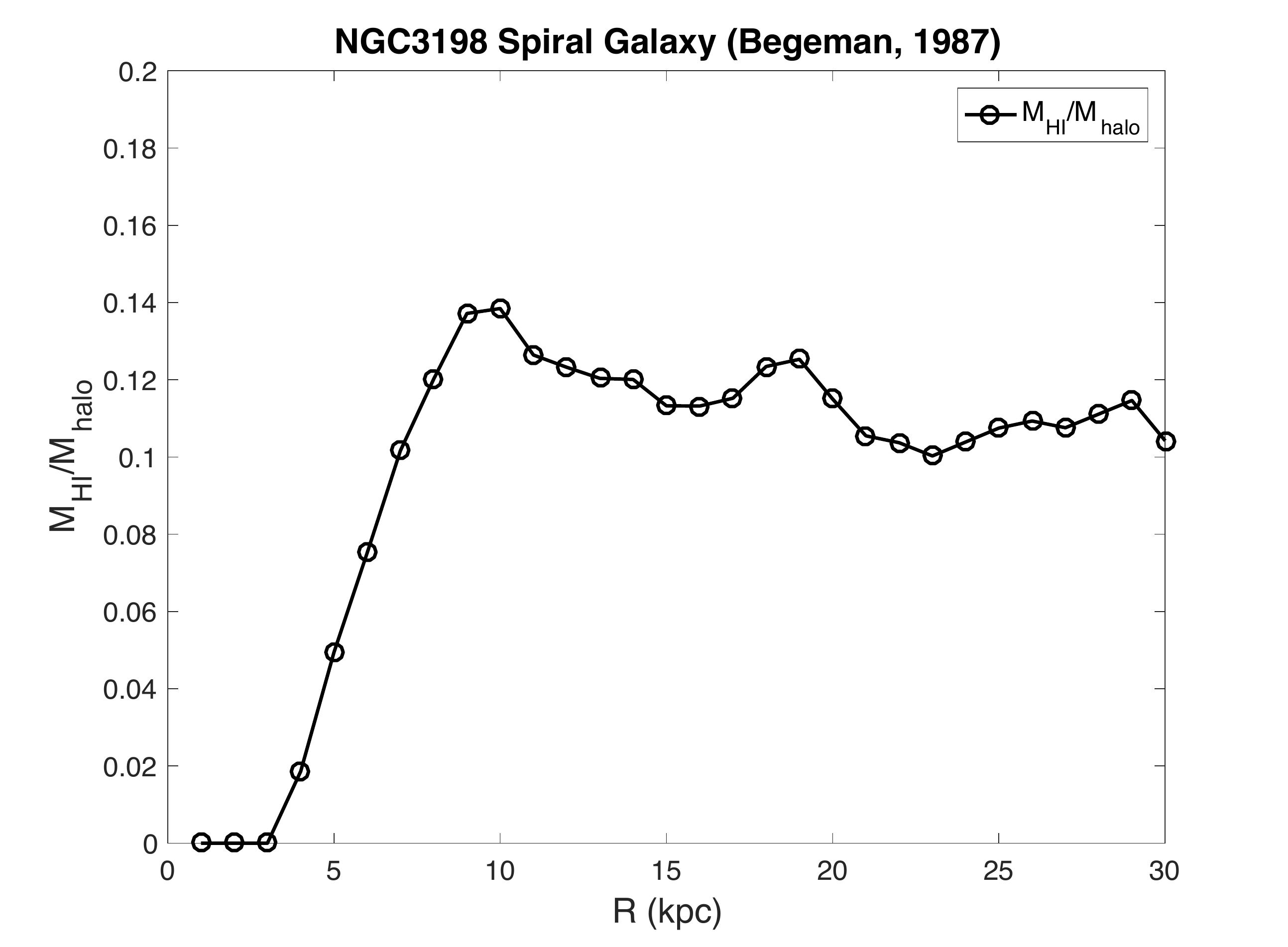}
         \caption{
(left) Begeman's (1987) cumulative radial mass distributions for the NGC3198 galactic mass contained in stars (M$_{HI}$ -- red symbols),  radioluminous $\tcmuboxdot$-HI gas clouds (blue symbols), and dark halo material that I propose to consist of of $\tcmuboxplus$- $\ce{H2}$ gas clouds (black symbols). Note that all stellar material is contained within a radius of $\sim \SI{8}{kpc}$, and that observed $\tcmuboxdot$-HI radioluminous matter and inferred $\tcmuboxplus$- $\ce{H2}$ dark matter increase roughly linearly at radial distances $>10-15$ kpc.
(right) Ratio of cumulative mass in $\tcmuboxdot$-HI gas clouds to dark matter mass in  $\tcmuboxplus$- $\ce{H2}$ gas clouds as a function of radial distance. There is a roughly uniform ratio of $12\%$ as much mass in radioluminous $\tcmuboxdot$-HI gas clouds as there is in the inferred trace-charged dark matter gas clouds of $\tcmuboxplus$- $\ce{H2}$. Data used are Begeman's (1987), shown in panel (A).
}
    \label{fig:Fig6ab} 

\end{figure}

If we compare this inferred dark-matter distribution (Figure 6) to Begemen's measured(\cite{bege1987,bege1989}) $\tcmuboxdot$-HI distribution (Figure 6), the $\tcmuboxdot$-HI fraction is a roughly uniform 12\% radial mass fraction of the dark matter necessary to explain the observed rotation curve (Figure 5) of $\tcmuboxdot$-HI matter in the distal portions of this galaxy.  Trace-charged dark matter appears to provide a simple explanation for the observed flat rotation curves of spiral galaxies. The hypothesis is compatible with current observation constraints that the dark-matter halo around spiral galaxies decays as $\rho (R) \propto R^{-2}$. (Note that the simple spherical approximations assumed to represent more probable ellipsoidal mass distributions do not affect the approximate form of these solutions, while only changing quantitative results by $\sim$20\% (cf. \cite{binn2007}, section 2.5)).

In the limit of a small $\tcmuboxdot$-HI gas fraction $\fboxdot$, small clouds of HI-gas are what rotate to produce the observed rotation curve, while their embedding cold $\tcmuboxplus$-$\ce{H2}$-gas clouds (Figure 4) provide the bulk of the mass that causes their flat rotation. The two different type of gas clouds experience differing electrogravitational forces which tend to make them rotate at different speeds, leading to a well-stirred $\tcmuboxdot$-HI distribution. Second order frictional (or `viscous') effects that transfer angular momentum between adjacent $\tcmuboxdot$-HI and $\tcmuboxplus$-$\ce{H2}$ clouds will tend to cause $\tcmuboxdot$-HI clouds to slow down and migrate inwards towards the center of the galaxy while the larger mass-fraction of $\tcmuboxplus$-$\ce{H2}$ clouds will be sped up and repelled outwards to an extent that depends on the effective viscosity and the relative mass-fractions of $\tcmuboxplus$ dark matter and $\tcmuboxdot$-HI matter. Qualitatively, the smaller mass-fraction $\tcmuboxdot$-HI matter would be influenced more. Distributions of  $\tcmuboxdot$-HI and $\tcmuboxplus$-$\ce{H2}$ are predicted to both fall off as $r^{-2}$. 

Observations of NGC3198 suggest that $\tcmuboxdot$-HI clouds have roughly $\sim 12\%$ of the mass needed to produce its flat $HI$-rotation curve, consistent with unseen $\tcmuboxplus$-$\ce{H2}$ forming $\sim 88 \%$ of the total mass in the distal regions of this galaxy. Each of their mass-distributions resembles that of the well-known Bonnor-Ebert isothermal sphere  (\cite{bonn1956,eber1955}) of equation (30), with the important distinction that dark-matter halos are not susceptible to gravitational collapse until their trace-charge is removed and expelled. Due to slow galactic rotation speeds, net magnetic forces within rotating $\tcmuboxplus$-$\ce{H2}$ clouds will typically be much smaller than Coulomb and gravitational forces (Figure 7), with the possible exception of the central region near a rotating ultramassive trace-charged black hole.

\begin{figure}
	\includegraphics[width=\columnwidth]{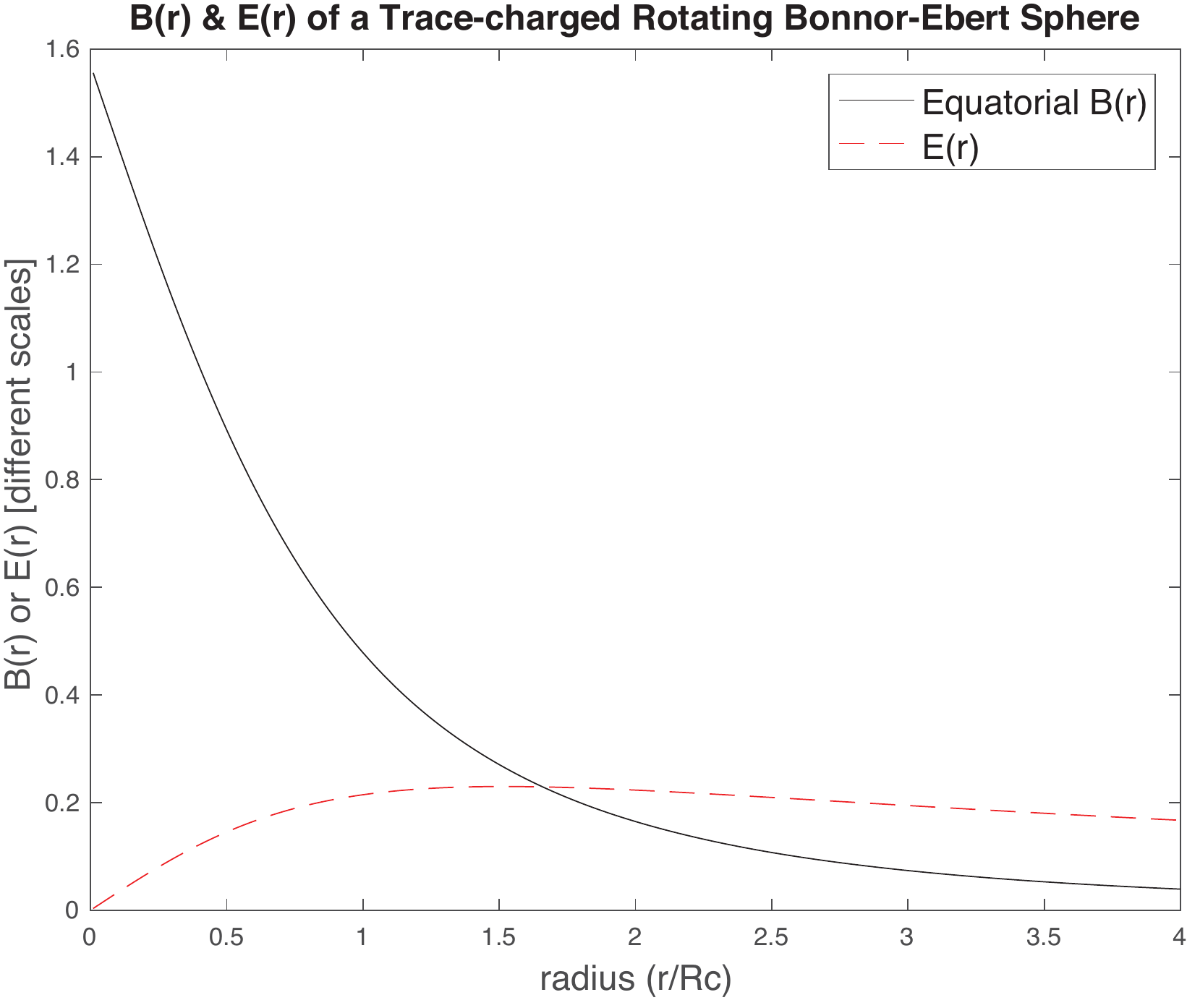}
         \caption{
Radial dependence of the electric and equatorial magnetic fields resulting from uniform circular rotation at speed $V_{ROT}$ of spherical shells of the Bonnor-Ebert spherical $\tcmuboxplus$- $\ce{H2}$ trace-charged halo mass distribution in Figure 6 (equation 29)  with charge density $\cboxplus \epsCG \rhoplus(R)$. Note that the equatorial magnetic field $B_{eq}(R)=(\mu_{0}\rho_{0}\cboxplus \epsCG V_{ROT} R_{C}/3)(\pi/2 + (R_{c}/R)^{3} \ln(1+(R/R_{C})^{2}) - \arctan (R/R_{C}) - (R_{C}/R))$ is of order $(V_{ROT}^{2}/c^{2})$ smaller than the electric field $E(R)=4 \pi \rho_{0} k_{E} \cboxplus \epsCG R_{C} (R_{C}/R) (1- (R_{C}/R)\arctan (R/R_{C}))$ at any given radius, this figure shows only how these fields decay radially from the Galaxy's center. Both fields extend well into intergalactic space.
}
    \label{fig:Fig7BE_BonnorEbertRotatingSphereEPS} 

\end{figure}

If this hypothesis is correct, then interactions between diffuse `uncharged' $\tcmuboxdot$-HI clouds and trace-charged $\tcmuboxplus$-$\ce{H2}$ clouds will clearly be an important factor shaping the dynamics of galaxies. Their differential rotation speeds could possibly be linked to the creation and maintenance of spiral arm structures and star formation within these dynamically-created features. Incorporation of the above two-gas theory into galaxy dynamics codes -- and simpler exploration of solutions to equations (16,17, and 19), with and without an appropriate intra-cloud `viscous friction' term -- is also likely to lead to other falsifiable implications of this hypothesis.

The inference from this interpretation of the observed dynamics and HI-gas distribution in NGC3198 is that unseen trace-charged dark matter contains of order $\sim 8$ times the mass of HI-radioluminous matter within the distal regions of spiral galaxies.  Can we detect a similar mass-ratio of cold $\tcmuboxplus$-$\ce{H2}$ matter to $\tcmuboxdot$-HI matter within the Milky Way?

\section{Speculation on charge segregation induced by this Universe's Big Bang genesis}

I hope the preceding sections have convinced you that the presence of trace-charged dark matter should be considered a serious candidate hypothesis to explain currently puzzling aspects of the solar wind, the origin of cosmic rays, the powersource of AGNs and pulsars, and the form of dark matter responsible for the observed apparent non-Keplerian rotation curves of distal HI-clouds in rotating spiral galaxies. Once we provisionally accept the possibility that regional trace-charge imbalances exist in the Universe at the level where their Coulomb charge effects are comparable to those of gravitational forces, a logical next question is `Why?'. To a degree, this question is a distraction from attempting to find tests and falsifications of the hypothesis itself. Cosmology is rife with fundamental assertions, for example the idea that charge must be exactly conserved during particle genesis (as, prior to the discovery of parity-violations and charge conjugation-parity violations, matter and anti-matter were thought to be), or the cosmological principle that the Universe is the same in all places and all directions. Next I will briefly outline my currently preferred speculations, in the hope that they provide help to others working to progress in this area. (Similar highly speculative exploration of the related question `Should charge always be exactly balanced in the cosmos even if the positive charge-carrying proton is much more massive than the negative charge-carrying electron' is what inspired the work described above.) The following speculations on the initial and early conditions within the Universe remain highly cosmopoetical, in large part because they lack guidance from a solution to Einstein's equations describing space-time within a black hole containing distributed matter.  

One line of speculation is that, just as the Universe is apparently dominated by matter over antimatter, presumed to be a consequence of a trace `CP violation' in the electroweak force, the Universe also formed with a slight excess of positive charge. (Alternatively, perhaps the trace asymmetry in the electroweak force is linked to the fact that the near-Solar environment is characterized by a trace excess of positive charge as discussed above?). I do not see how to test or further refine this hypothesis, except for its potential links to apparent `dark energy' as discussed below. That does not make it untrue, just undesirable in terms of making further progress. 

The speculation that I have played with, to the point of constructing toy process-models of the post Big-Bang expansion of the Universe, is that the Big Bang itself was the origin of mass-linked charge segregation in the earliest Universe. The mechanism is related to how a supernova could transform nearly uncharged stellar matter into a trace-charged electron rich periphery of a supernova explosion and a proton-rich neutron star at its core. It starts from the assumption that the Universe is and was matter dominated, and that the positive charge-carrying proton is $\sim$1836 times more massive than the negative charge-carrying electron. It also imagines the Universe to be the consequence of a Maxwell-Boltzmann-like (or relativistic Maxwell-Juttner-like) `explosion' from a localized Giant black hole, or two colliding Giant black holes (see Figure 8). The two initial Giant black holes that collided within the Cosmos they were embedded within could already have had initial net charge imbalances as long as their original charge/mass ratios were less than $\epsCG$.

\begin{figure}
	\includegraphics[width=10cm]{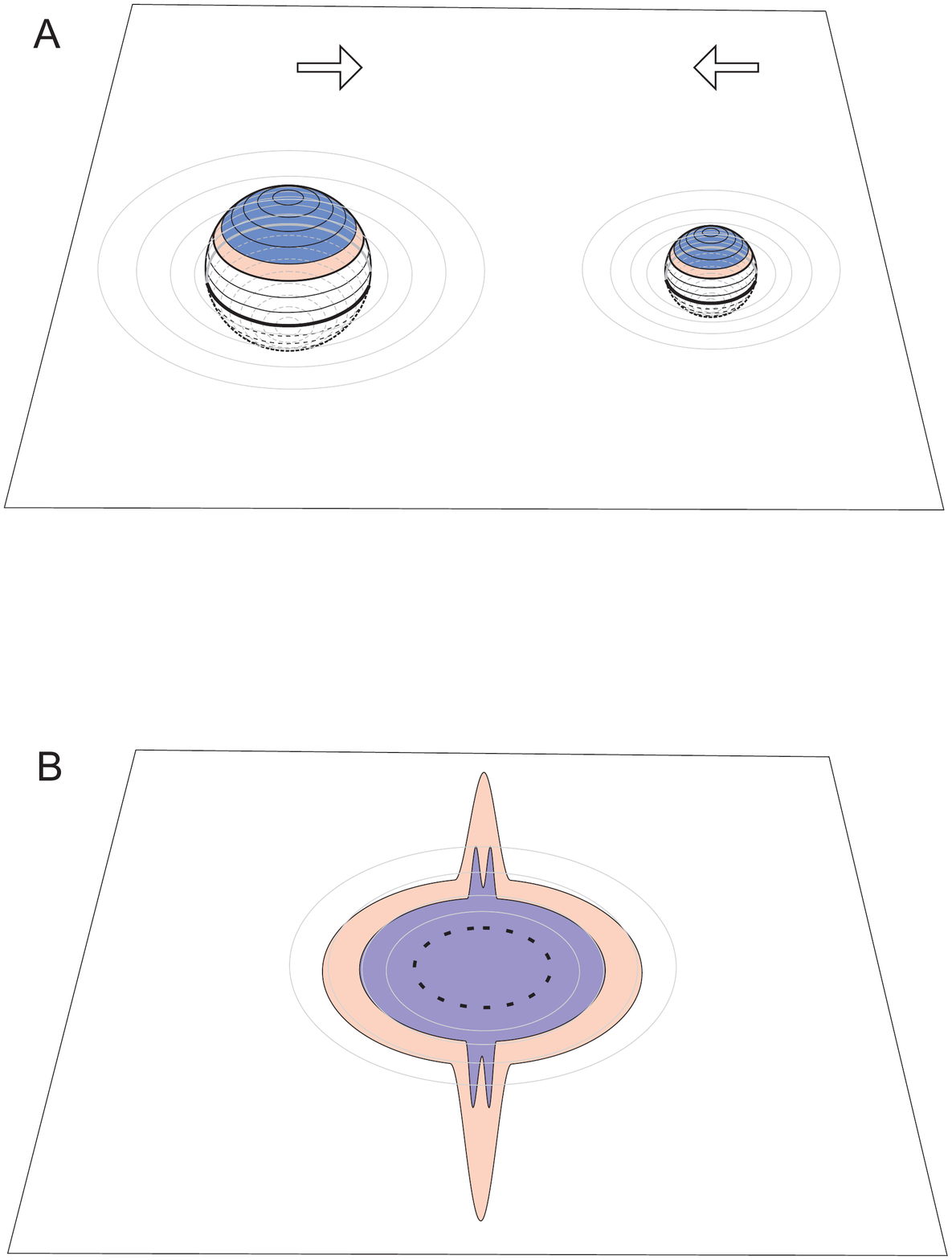}
         \caption{
Two-dimensional space-time sketch of a Big Bang scenario analogous to the one explored in this study. I speculate that  the collision of two Universe-mass-like black-holes could trigger this event, each black hole having its matter concentrated into a concentrated lump around its center in which charge segregation already exists, with each having an electron-rich rind (pink) and an electron-poor core (blue-see text). The collision would lead to a significant fraction of the kinetic energy of the colliding black holes being transformed into kinetic energy of escaping `jets' composed preferentially of  material from the outermost electron-rich rinds at the site of the initial collision. (see text). This preferential loss of negative charge triggers a Coulomb explosion within the remaining matter. Once the matter has expanded so that its density is less than the Schwarzschild density, photons will escape, leading to an irreversible Big Bang-like event. In these cartoons, the external Schwarzschild radii are shown by the heavy black circles. Within each black hole, space-time is cartooned as a closed sphere (the shape should have different curvature within the matter-containing region). External space-time is cartooned as being flat except in the `vicinity' of each Giant black hole, where idealized potentials of gravitational wells are sketched as grey lines.
}
    \label{fig:Fig8BigBangJettingSketch} 

\end{figure}

Even without an appropriate solution for the space-time fabric within a black hole containing distributed matter and charge, we can use conventional physics to (mis?-)imagine some of the initial conditions. I consider the following points as guideposts for speculation.

(A) The mean density of a black hole that contains the amount of matter in the Universe is very low.
If we estimate the mean density within a black hole as its mass (including the energy-mass carried by photons within the black hole) divided by its external Schwarzschild volume $4 \pi R_{S}^3 / 3$  (where $R_{S}= 2GM/c^2$), we obtain

\begin{equation}\label{E:Eq32}
\rho^{}_{UBH} = \frac {3 M}  {4 \pi R_{S}^{3}} = \frac {3 c^{6}}  {32 \pi M_{UBH}^{2} G^{3}} 
\end{equation}

\noindent
For an initial Universe mass of $\SI{1e51}{kg}$ (note this is 3-orders of magnitude smaller than recent estimates of matter + dark-matter based on the standard cosmology paradigm), the corresponding black hole radius would be $\SI{1.5e24}{m}$  ($\SI{48}{MPc}$) and its mean density would be $\SI{7.3e-23}{kgm^{-3}}$ (e.g. $\sim$2 orders of magnitude lower than the density of the present-day interstellar medium). For a ten-fold increase in mass, the mean density decreases by a factor of 100, and the radius by a factor of 10.  As large black holes have negligible Hawking radiation they cannot effectively lose their internal thermal energy. This means they are likely to be hot, as all of the kinetic energy and a significant fraction of the radiation emitted from infalling material near their event horizon will be captured within the black hole to contribute to its thermal energy. 

(B) The thermodynamic equilibrium mass-energy associated with radiation within the black hole is significant, even dominant at high temperatures. Olbers paradox applies inside black holes, they should have an equilibrium `bright sky' inside.  The lowest mass-energy + energy state will often be one in which temperature, hence radiation energy tends to be minimized, even at the expense of matter aggregating into clumped matter configurations with lower near-zero binding energies.  

(C) If we do not include gravitational red-shift effects within the black hole, then the photon mass-energy and radiation pressure will quickly dominate the effects of any matter within the black hole. For example, even for an internal black-hole temperature of $\SI{400} {K}$, the radiation pressure $P_{rad} = aT^{4}/3$ leads to a radiation mass-energy density $P_{rad}/c^{2} = aT^{4}/(3c^2) = \SI {7.2e-23} {kgm^{-3}}$ with $a=8 \pi^{5} k_{B}^{4} / (15 h^{3}c^{3}) = \SI {7.565e-16} {JK^{-4}m^{-3}}$. This radiation mass-energy density for a cool temperature of $\SI{400} {K}$ is already the same order of magnitude as our previous estimate for the mean density of a black hole of mass $\SI{1e51}{kg}$ -- and the black hole's internal radiation mass-energy depends on $T^{4}$. 

(D) A single aggregation of matter will be the lowest net energy state, therefore the one favored for internal equilibrium.  Radiation mass-energy is minimized by concentrating matter within the black hole so that, by its gravitational red-shift effect, it can reduce the average  mass-energy of the photons that are in a equilibrium distribution within the matter-free portions of the black hole. The consequence is that, like stars with relativistic radiation pressures, the resulting single large clump of matter within a black hole may only be weakly bound and susceptible to explosive disassembly should it be disturbed during a large collisional or accretional event. (In other words, any increase in its matter's binding energy would lead to a higher net temperature within the black hole, hence a higher equilibrium radiation mass-energy. This leads to an energetic preference for matter to be in a highly unstable state as long as it has the lowest stable binding-energy equilibrium.)

(E) A state that can minimize the binding energy of matter within the black hole is one in which trace-charge imbalances slightly overcompensate the gravitational attraction of the matter ($\tcmuboxplus$ with $\cboxplus > 1$), with the electrical attraction of an electron-rich rind and  atmosphere providing the minimum positive binding energy for the clumped matter, and radiation pressure, including its gravitational red-shift effects, adding additional radiation support and a possible equilibrium thermal gradient to this matter.

To approximately determine this configuration I consider the central matter of mass $M$ to have uniform density $\rho$ and charge density $\cboxplus \epsCG \rho$, with the corresponding electron-rich rind at radius $r_{0}$ having a net charge $Q^{-} = -\cboxplus \epsCG M$ (Note this analysis can be easily extended to the case where the black hole's net trace charge is positive, as I do next when considering the approximate charge-configuration within a black hole with a trace-positive binding energy. This analysis also neglects the mass-energy effects of electric fields.).  This matter and charge configuration would have a `classical' internal pressure distribution 

\begin{equation}\label{E:Eq33}
P(r)= \frac {2} {3} \pi \rho^{2} G \left( r_{0}^{2} \left( 1 - \frac {\cboxplussquared} {3} \right) + r^{2} \left( \cboxplussquared -1 \right) \right), 
\end{equation}

\noindent
with a corresponding $p-V$ energy 

\begin{equation}\label{E:Eq34}
\frac {2 \pi \rho^{2} G r_{0}^{5} \lbrace 2 + \frac {4} {3} \cboxplussquared \rbrace} {45}  
\end{equation}

\noindent
and $p-V$ mass-energy 

\begin{equation}\label{E:Eq35}
\frac {2 \pi \rho^{2} G r_{0}^{5} \lbrace 2 + \frac {4} {3} \cboxplussquared \rbrace} {45c^{2}} . 
\end{equation}

\noindent
The electrogravitational binding energy of the central trace-positively charged region is $3(\cboxplussquared -1)GM^{2}/(5r_{0})$. If we neglect its $\sim$1836 times smaller gravitational attraction, we can approximate the rind's binding energy as the Coulomb binding energy of a pure-electron rind of charge $fQ^{-}$, which is $-k_{E}Q^{2}f(1-f/2)/r_{0} = -\cboxplussquared GM^{2}f(1-f/2)/r_{0}$. Combining these terms, the net binding energy is given by

\begin{equation}\label{E:Eq36}
U =  \frac {3(\cboxplussquared -1)GM^{2}} {5r_{0}} - \frac {\cboxplussquared GM^{2}f(1-f/2)} {r_{0}}
\end{equation}

\noindent 
which implies a zero net binding energy when $U=0$, e.g.  when

\begin{equation}\label{E:Eq37}
\cboxplus =  \left(1- \frac {5f(1-f/2)} {3} \right)^{-1/2}
\end{equation}

\noindent
When $f=1$ (e.g. no net charge contained within the black hole), this happens for $\cboxplus = \sqrt{6} \simeq 2.45$, when $f=0$ (e.g. a maximally trace-charged black hole), $\cboxplus = 1$, and for a non-maximally trace-charged black hole $1 \le \cboxplus \le \sqrt{6}$. 

For the purposes of this discussion, the key point is that the equilibrium mass-distribution within a Giant black hole could well be almost critically unstable, with its electrons already concentrated towards its outer surface.  The electrons lower mass with respect to protons would also lead to them being concentrated within its `atmospheric' outermost rind.

Now imagine what would happen if two such Giant black holes collide. (Another conceivable tipping point could be the grazing collision of two ultramassive black holes, with one robbing negative charge from the other, triggering a similar explosive event). The kinetic energy of their collision would be initially transferred into their collision interface. I envision this would lead to mass-jetting (Figure 8) like that which happens in any large impact.  If this mass-jetting preferentially ejects the electron-rich rind of near-surface material, enough charged matter could leave the transient black-hole assemblage for it to then undergo a Coulomb explosion (cf. \cite{pant2012}), which will happen when $|\tcmuCG|>\epsCG$ if the electron-rich rind no longer provides an adequate restoring pull.

Retarded-time effects appear to make the receding electron-rich rind only feel the gravitational attraction of the central matter, not its net Coulomb charge (see Figure 9). This effect makes the electron-rich rind's escape much easier than would be predicted assuming instantaneous propagation of electric fields. 

\begin{figure}
	\includegraphics[width=9cm]{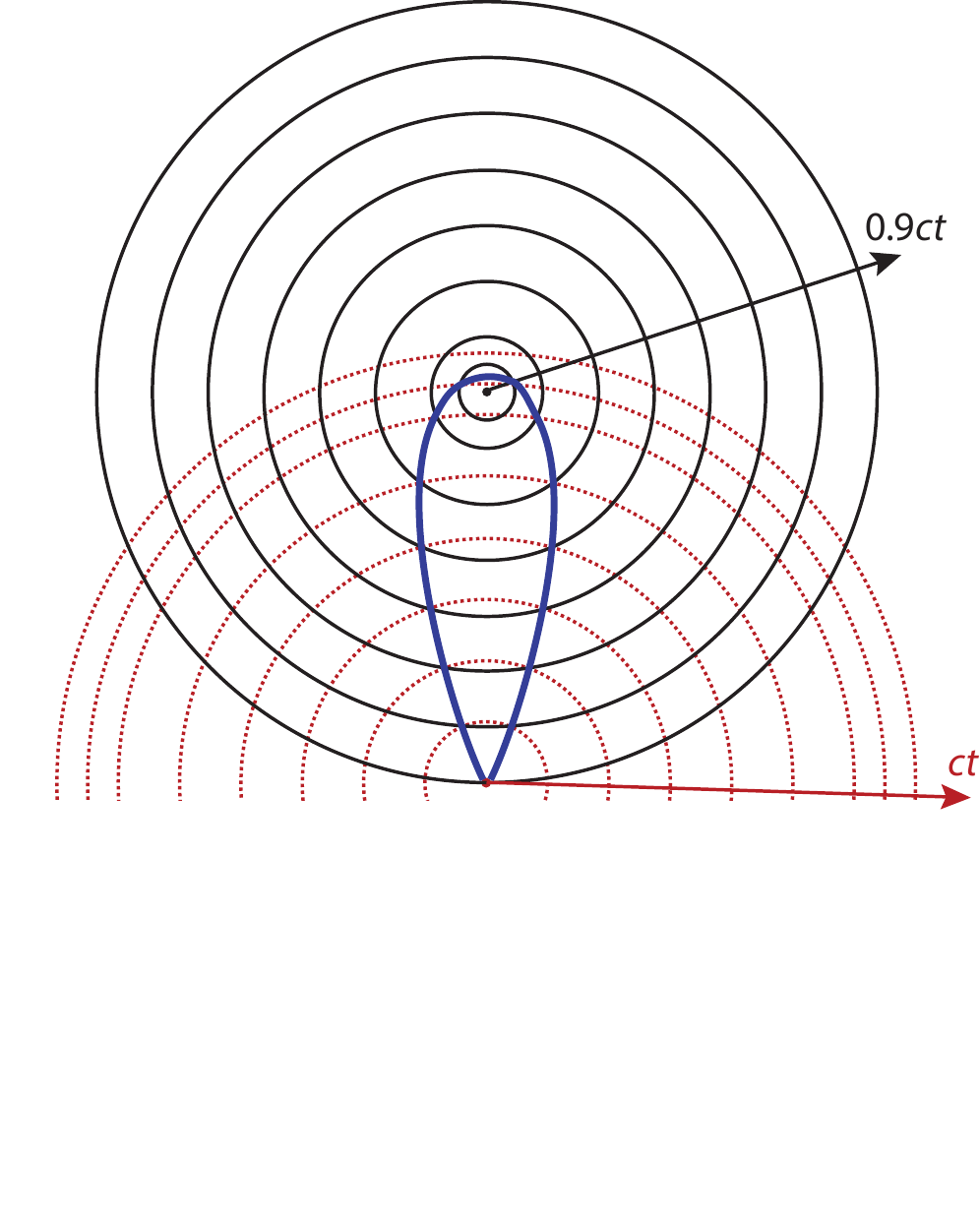}
          \caption{
Sketch showing the apparent location of an expanding ring or shell of net negative charge, when this ring is expanding at $0.9c$. Retarded-time effects become important at  relativistic expansion speeds.  From the perspective of the point of the outermost ring of expanding charge that is marked by a star, the apparent location of this ring looks like the blue teardrop-shaped curve. This teardrop shows the retarded-time arrival of the electric field for all points on the outermost expanding ring of charge. It is given by the intersection of the curves for electric field propagation at $c$ and outward ring expansion at $0.9c$. The mean apparent distance of net negative charge is always equal to, or closer than the mean distance of the slower expanding core with net positive charge, implying that only the gravitational attraction of the central mass, not its Coulomb charge, has a retarding influence on the outward expansion of a relativistic receding ring or shell of charge.  (Note that as the outward expansion rate  of the ring/shell of net negative charge approaches $c$, the teardrop will compress into a compact sphere about the original center of mass with a single `ray'(ring) or `narrow cone'(expanding spherical shell) pointing in the direction of the observation point. In this case the observation point 'sees' the ring/shell at a retarded time that is still before the onset of the Coulomb explosion.
}
    \label{fig:Fig9BigBangMatterDensity} 

\end{figure}

 I imagine the initial particle velocity distribution can be approximated as a quasi-equilibrium Maxwell-Boltzmann distribution. Once the black hole's internal density is reduced to the point where its space-time opens to the Cosmos, photons within the black hole will escape, leading to the Big Bang proper.  Once photons start to escape in significant amounts, the loss of exterior radiation pressure on the central matter will further contribute to its explosive fragmentation and dispersal in a bright Big Bang. Internal scattering from the latter, appropriately gravitationally red-shifted (\cite{bond1999,Bonn1975}), could be the source of today's CMB blackbody photon distribution.

Note that smaller ultramassive black hole collisions or large black hole accretion events could be associated with similar explosive failure if the black hole is hot and massive enough to have this barely stable internal distribution of matter, charge, and energy. In general there might be a pair of photon flashes, an initial flash associated with the ejection of the initial ultrarelativistic jets of negatively charged matter beyond a still-present event horizon, and a second flash associated with massive photon escape as the event-horizon itself dissolves. 

(F) If the Universe started from the disintegration of matter that had an initial Maxwell-Juttner-like velocity distribution, it would have a M-B-like velocity distribution within its central, slower moving $\rho (v) \propto v^{2}$ portion. Continued outward expansion of this portion of the velocity distribution has the interesting property that it would generate the local appearance of a distribution with uniform density, uniformly expanding matter -- exactly the distribution needed to explain the presence of a `Hubble Constant' for the apparent uniform expansion of local regions of our Universe. 

\begin{figure}
	\includegraphics[width=10cm]{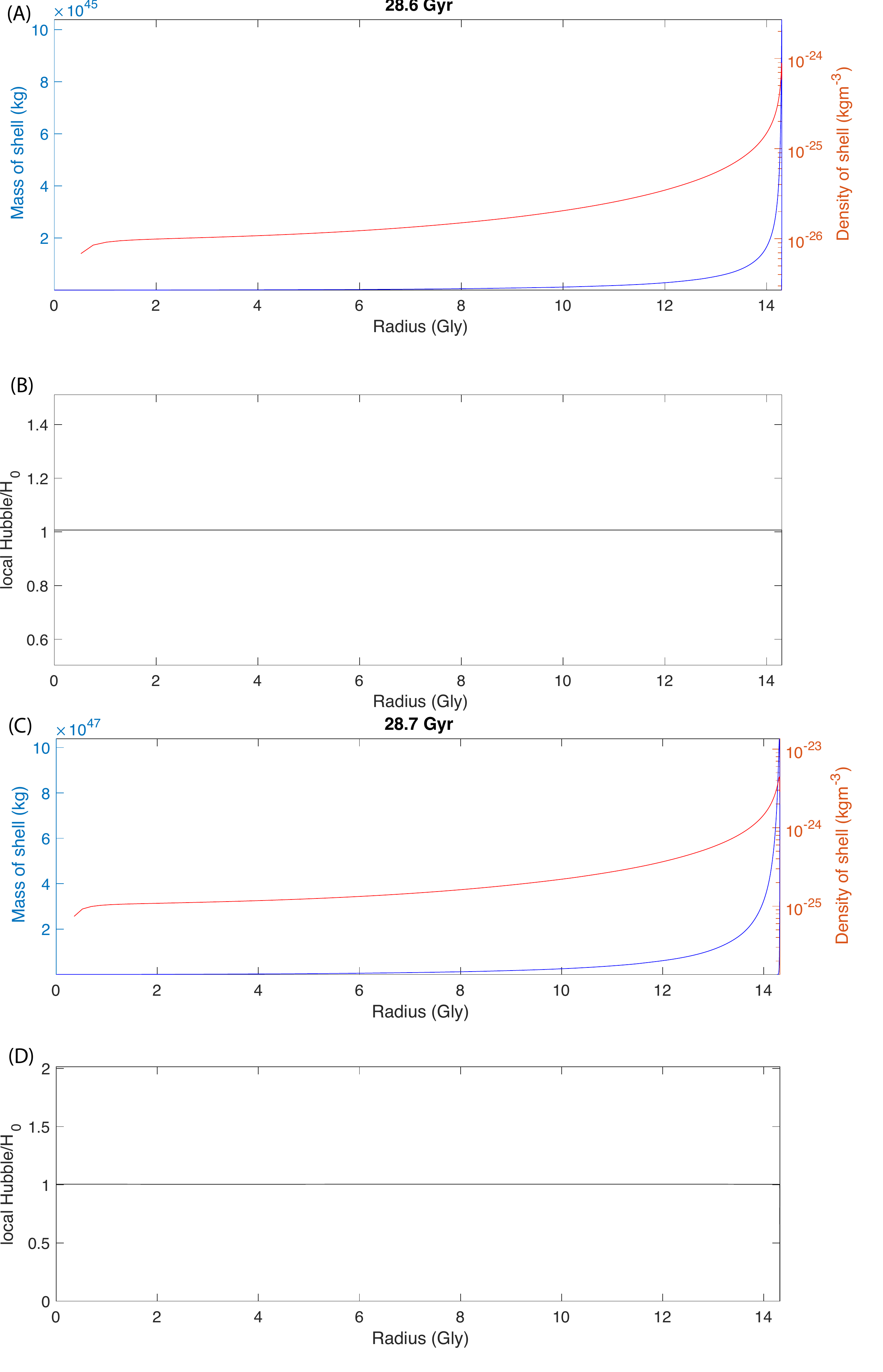}
          \caption{
(A) Radial Mass (blue line) and Density (red line) distributions at time 28.6 Gyr for a Coulomb-explosion generated Universe with an initial mass of $\SI{1e49}{kg}$, initial temperature of $\SI{4e10}{K}$ and initial central $\cboxplus = 1.2$. (The explosion is only a weak function of initial temperature.) (B) Radial Hubble constant distribution at time  28.6 Gyr for this toy-model Universe. At this stage this model run would have the Universe's observed Hubble constant and a density similar to the critical density $\rho_{crit} = 3H_{0}^{2}/8\pi G =\num{8.9e-27}$ for flat Euclidian space-time.
C) Radial Mass (blue line) and Density (red line) distributions at time 28.7 Gyr for a Coulomb-explosion generated Universe with an initial mass of $\SI{1e51}{kg}$, initial temperature of $\SI{4e10}{K}$ and initial central $\cboxplus = 1.02$.  (D) Radial Hubble constant distribution at time  28.7 Gyr for this toy-model Universe. At this stage this model run would have the Universe's observed Hubble constant.
}
    \label{fig:Fig10ABCDNEW} 

\end{figure}

This predicted behavior is evident in a 1-D spherical toy-model that treats only the post-disintegration expansion of this Big Bang scenario. (The Appendix gives more detail on the numerical treatment in this model.)  The initial condition for each experiment is a compact region within a Universe-sized Giant black hole that contains an equilibrium Maxwell-Boltzmann distribution (eq. 5) of already-generated protons and electrons. This region has been perturbed by the ultrarelativistic jetting of a small fraction of peripheral electron-rich 'impact jets' associated with the first phase of the Big Bang (Figure 8). The resulting Coulomb-dominated expansion of internal positively trace-charged spherical shells of matter leads to the density-time distributions of radially expanding shells of matter shown in Figure 10.  The evident `uniform everywhere' appearance within the expanding shells of matter is due to two factors. First the initial distribution of matter has a significant central $\rho (v) \propto v^{2}$ portion whose expansion would naturally lead to a central region with a uniform distribution of expanding matter. Second, when Coulomb repulsive electrostatic forces are larger than gravitational attractions ($\cboxplus > 1$),  central shells of expanding matter can overtake outer shells due to Coulomb-repulsion-accelerated internal expansion. This effect also leads to a uniform distribution of matter. Resulting local Hubble constants are even more uniform than the matter distribution. Figure 10 shows two snaphots of the matter distribution and local Hubble constants for expansion from model Universes. After the initial phase, the density of expanding matter is quasi-uniform with radius. Figure 11 shows the time variation of the `Hubble constant' in the toy-model simulation shown in Figure 10AB. At all times, the Hubble constant is quasi-uniform with radius (see Fig. 10A,10C). Due to initial strong Coulomb-repulsion effects, the Hubble constant has an apparent `inflation-like' early phase (Fig. 11), in particular in the earliest time-period when inflation occurred within the mass's Schwartzschild radius. During this early phase, could the high internal photon density within the still-present black hole  help to keep the expanding mass near a state of thermal equilibrium, resulting in a nearly uniform temperature distribution preserved within CMB photons? 

\begin{figure}
	\includegraphics[width=14cm]{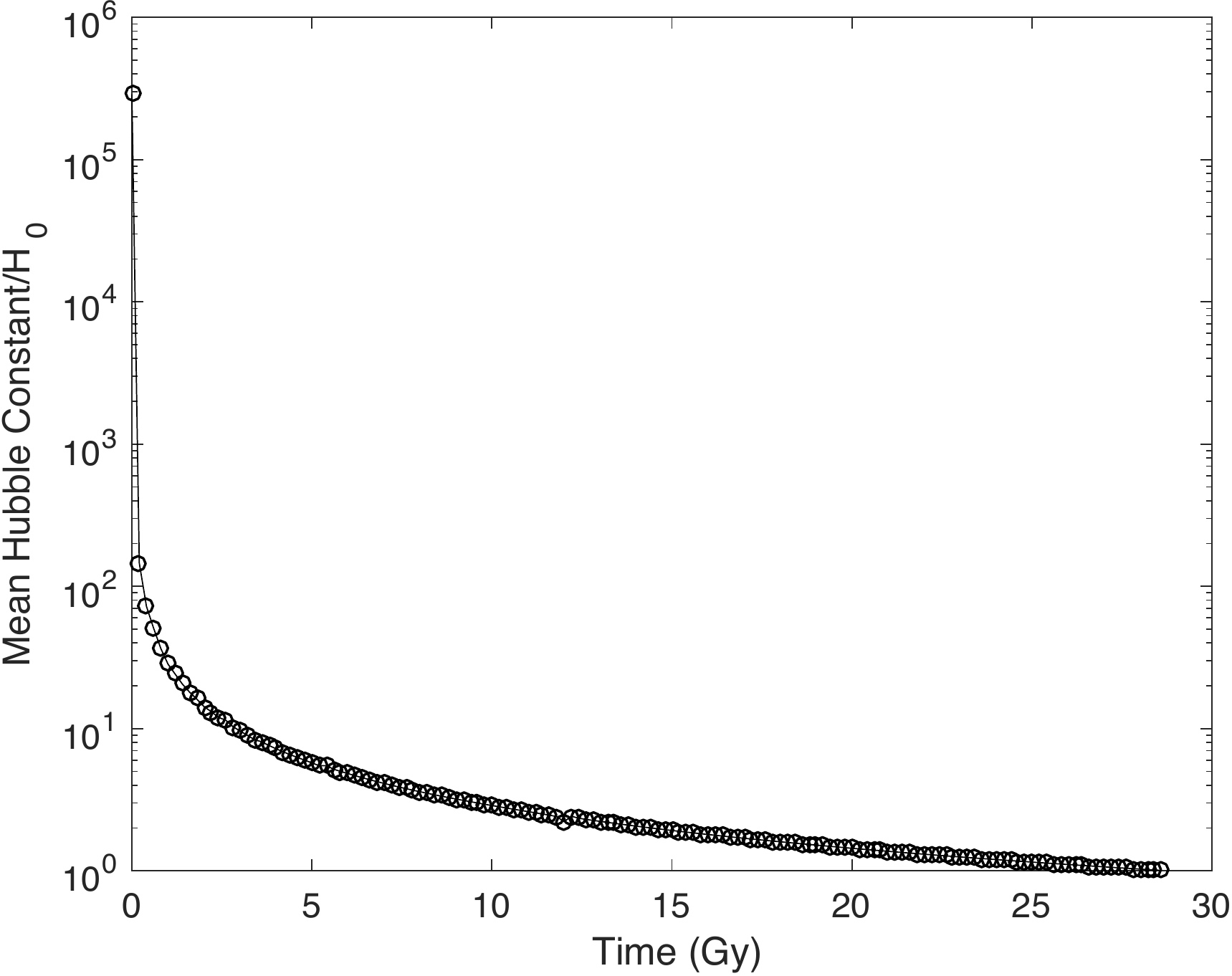}
         \caption{
Mean Hubble constant as a function of time during expansion from a Maxwell-Boltzmann Coulomb explosion with initial mass M$=\SI{1e49}{kg}$, temperature T$=\SI{4e10}{K}$ and central charge-to-mass ratio $\cboxplus=1.2$, for the toy-model experiment shown in Figure 10AB.
}
    \label{fig:Fig11HubbleVsTime} 

\end{figure}

With this toy model scenario, it is straightforward to generate nearly uniform mass distributions of expanding trace-charged matter with $\cboxplus \ge 1$ for initial masses of order $\num{1e49}-\SI{1e53}{kg}$. With lower mass Universes ($\sim \SI{1e49}{kg}$) it is easier to generate low present-day local mean densities of $\SI{1e-26}{kg}$, if this (low) density is needed to produce the observed curvature of space-time. In general, two Hubble times (e.g. $\sim 28 Ga$) of expansion is needed to generate a current expansion rate equal to the measured Hubble constant.

\section{First-order Cosmological Implications}

The trace-charged dark-matter scenario appears to have the potential to provide simple, straightforward interpretations for several first-order cosmological observations and puzzles such as: 

\noindent
(1) the ``cosmological principle'' of `uniform everywhere behaviour' is a byproduct of a Coulomb explosion generating the mass-distribution in our region of the Cosmos. The Galaxy happens to lie within outward expanding matter from the $\rho (v) \propto v^{2}$ portion of the Big Bang's initial quasi-equilibrium Maxwell-Boltzmann velocity distribution. This type of Big Bang should have a definite origin in space and time within the Cosmos. The dipole term in CMB radiation could be linked to our true motion with respect to this frame.

\noindent
(2) The CMB is due to scattering and the gravitational red-shift of photons associated with the initial breakout and `event-horizon dissolution' of the Big Bang's proto black hole when its mean density decreased below the Schwarzschild density for this mass. Brandt (\cite{bran2008}, section 6.3) indicates how scattering from uniformly receding matter in the Universe is a `classical' mechanism to further red-shift the CMB from this neo-classical Big Bang. The resulting CMB would still be a key indicator that the expansion of the Universe began with a Big Bang, but not as conventionally envisaged based on 1-D kinematic solutions to the Einstein field equations.

\noindent
(3) The 'cosmological constant' ($\Gamma$) could well be zero.  The assertion of a non-zero  $\Gamma$ has been a significant recent change in our cosmological interpretation of the expanding Universe, motivated by observations that the Universe appears to have a `negative gravitational pressure' that is counteracting the effects of normal massive gravitation. Instead, in the trace-charged Big Bang hypothesis, the apparent repulsive effect is due to familiar electrostatic repulsive forces between regions of $\tcmuboxplus$. There appears to be no need for the cosmological constant to be anything but zero, the simplest value consistent with local gravitational measurements, observations on the dynamics of the solar system, and the equations of general relativity.

\noindent
(4) The apparent 'accelerating' expansion rate of the Universe. Measurements appear to suggest that the rate of expansion in the recent Universe has accelerated relative to the rate implied by distance-red shift observations on early (old) quasars. (If the strong red-shifts of quasars are actually gravitational red-shifts of exploding black holes as speculated soon after their discovery (cf. \cite{bond1999,Bonn1975}), then this observation may be a misinterpretation). 
If the preferred state of trace-charged matter inside Giant hot black holes is a spherical core of $\tcmuboxplus$ surrounded by an outermost rind of electron-rich $\tcmuboxminus$, and the net charge of the black hole is positive, then $\cboxplus>1$ in initial $\tcmuboxplus$. It seems quite possible that the central region of the expanding Big Bang could continue to have a higher positive trace-charge than more outward regions, with the outermost regions that started with a net negative trace-charge ultimately reversing direction,  recombining with and reducing the net charge of outer regions of $\tcmuboxplus$ matter during their inward fall towards the center of the initial explosion. This radial charge distribution with $\cboxplus$ decreasing radially outwards would lead to Coulomb repulsion forces that create the apparent effect of `Dark Energy' acceleration of expansion' within the central, more strongly trace-charged regions of the Big Bang. However, in this scenario, closer, more recently emitting regions only expand faster because they also contain more strongly trace-charged dark matter, not because the overall expansion rate of the Universe is actually increasing. Perhaps we are currently misinterpreting spatial expansion variations within the Universe as evidence for temporal `Dark-Energy-linked' $\Lambda$CDM accelerating expansion of the Universe.

\noindent
(5) Exotic $\Lambda$CDM forms of matter are not needed to explain current observations.

\section{Other Questions for Further Study}

The eventual strength of a conjecture lies as much in its ability to to generate new testable questions as it does in its ability to explain the observations and anomalies that led to its creation. Many questions were raised and not answered in the earlier sections of this paper. Here is a short list of other speculative questions that might lead to future tests:

\subsection{Sol-centric}

\noindent
(1) Can we observe the predicted minute net positive charge of the Sun? Is this a possible source of the slight anomalies in the outward acceleration of Voyager I?

\subsection{Stellar Dynamics and Star Formation}

\noindent
(2) What is the equilibrium charge/mass profile between a star and its surrounding halo of $\tcmuboxplus$? Does this have a weak dependence on the mass and temperature of its chromosphere?

\noindent
(3) Does $\tcmuboxplus$ freeze stellar coalescence and birth in regions where $\lvert \cboxplus \rvert > 1$? Is there accelerated stellar evolution in regions where $\lvert \cboxplus \rvert << 1$?

\noindent
(4) How is trace-charge expelled as a star coalesces, warms, and ignites? How does the expulsion of trace-charge influence the ignition and evolution of nearby stars in regions where stars are born? Is this the mechanism by which star formation shapes later star formation?

\noindent
(5) How does the merging of $\tcmuboxplus$ and $\tcmuboxminus$ take place? (Both electromagnetic and gravitational forces are likely to shape this process.) 

\noindent
(6) Effects of $\tcmuboxplus$ on black hole dynamics and accretion. How much net trace charge can an accreting ultramassive black hole usually accumulate?

\subsection{Galactic Dynamics and Star-Galaxy/Black Hole-Galaxy Interactions}

\noindent
(7) Do (differential) $\tcmuboxdot$ and $\tcmuboxplus$ rotations within a galaxy shape the stirring and mixing of $\tcmuboxdot$ and $\tcmuboxplus$ matter, thereby shaping both galaxy structures, e.g. spiral arm evolution, and the distribution of star-forming conditions? 

\noindent
(8) Do the magnetic fields associated with the rotation of $\tcmuboxplus$ in galaxies influence the galaxies' structure and evolution? 

\noindent
(9)  A rotating galaxy's intrinsic magnetic field will tend to funnel accretionary charged material along magnetic field lines into its axial/`polar' regions. Is this seen?

\noindent
(10)  Do galaxies interact magnetically with neighboring intergalactic $\tcmuboxplus$-material? 

\noindent
(11)  Do systems of galaxies share angular rotation characteristics due to the ultraweak magnetic interactions between their  $\tcmuboxplus$? 

\noindent
(12)  Are quasars a scaled-up, AGN-powered, version of pulsar-generated radiation? 

\subsection{Big Bang Dynamics}

(13) Opposite to electrons, high mass He nuclei would have slower initial speeds and therefore be preferentially retained in the central regions of a Big Bang with respect to the distal regions. Is this seen?

\noindent
(14) Is the net charge of the Universe non-zero? If so, could this alternative lead to visible astronomical differences?

\section{Summary Perspective}

The  trace-charged dark-matter hypothesis for the evolution of the Big Bang and our Universe provides a simple cosmology with well-understood materials that can can explain the origin of current major enigmas such as the apparent non-Keplerian rotation curves of distal HI-gas in spiral galaxies,  an energy source for high-energy axial jetting from AGNs, the strong positive-charge imbalance of cosmic rays, and  pulsar behaviour of neutron stars.  In this cosmology there is neither a need for  `dark energy' nor for exotic WIMP forms of baryonic or non-baryonic  $\Lambda$CDM matter. I hope that this hypothesis will help the cosmological community to explore fruitful observational and theoretical questions that lead to productive new insights into the nature of the Universe and Cosmos.

\section{Acknowledgements}

The initial idea for this study, and roughly a third of the work was done while I was working at Cornell, including
a productive summer visit to LMU (Munich). The main work, calculations, and writing of this paper was done during six very productive summer weeks in 2016. I thank Giovanna Vannucchi, Enzo Vannucchi, Anna Luna Morgan, and Paola Vannucchi
for creating a wonderful summer environment for focussed thinking and writing, and Dave Mattey for giving me the flexibility to focus on my research during this summer. Thanks also to W. Jason Morgan for helpful discussions throughout the genesis of these ideas, and also for help (via Harvard's incredible library) in finding the most inaccessible reference material. This work was supported by a Wolfson Merit Research Award.

\appendix

\section{Appendix: Big Bang toy model}

The St\"{o}rmer-Verlet algorithm (\cite{hair2003}) is used to solve the relativistic equation of motion 

\begin{equation}\label{E:EqA1}
F_{j} = \gamma^{3} m_{j} \frac{dv_{j}}{dt} 
\end{equation}

\noindent
for the radial expansion $v_{j}(r)$ of the $j^{th}$ trace-charged spherical shell of matter with mass $m_{j}$ experiencing a force $F_{j}$. The initial velocity-mass distribution is given by a Maxwell-Boltzmann distribution (eqn. 5) for protons at temperature $T_{0}$. The force $F_{j}$ on mass-shell $j$ is given by

\begin{align}\label{E:EqA2}
F_{j}(r) = &G \bigg(  \Big(  \sum_{i \le j}^{} (\cboxplus)_{i}m_{i}  (\cboxplus)_{j}  - \sum_{i \le j}^{} m_{i} \Big)  \\ 
 &- 0.5m_{j} \left( (\cboxplus)_{j}^{2} -1 \right) \bigg)
\notag
\end{align}

\noindent
Each toy model experiment is started with a `coasting' timestep using the Maxwell-Boltzmann velocity-mass distribution (eqn. 5) for initial velocities and radial distances from a central point. This initial timestep is chosen to bring the outermost radius of expanding material to the characteristic radius for the initial clumped matter (see eqn. A3 below, of order 1ly). After this initial timestep, the timestep is changed every 10 iterations to the minimum of $0.1a_{i}/v_{i}$ for all mass-shells $i$ at that timestep, or 0.25Ga.

Early expansion has an `inflation-like' phase driven by internal Coulomb repulsion after the initial removal and relativistic jet-ejection of a fraction of the outer rind material containing a net negative charge $Q^{-}= -\cboxplus \epsCG M$, where $M$ is the initial mass of the Universe. See discussion around equations (33-37) for more details of this initial mass and charge distribution.

The initial ejection velocity $\gamma_{0}$ of the impact jet $Q^{-}$ that initiates the Coulomb explosion assumes that a significant fraction $(\gtrsim 10\%)$ of the kinetic energy of clumped matter within the colliding Giant black holes goes into the ejected material in the impact-jet. The energy transferred into the impact jet is estimated to be the difference in (gravitational) binding energy of two uniform density spheres with matter of mass $M/2$ and density $\rho_{0}$, and a post-collision sphere of mass $M$ and density $\rho_{0}$. This estimated kinetic energy is $37\%$ of the gravitational binding energy of a uniform density sphere of mass $M$.  Each Giant black hole is assumed to have at its core clumped matter of quasi-uniform density $\rho_{0}$ with a radius

\begin{equation}\label{E:EqA3}
R_{0} = \sqrt[3]{\frac{3M_{0}}{4\pi \rho_{0}}}
\end{equation}

\noindent
and potential energy 
\begin{equation}\label{E:EqA4}
U = \frac{\sqrt[\leftroot{2} \uproot{2} 3]{36\pi} }{5} GM_{0}^{5/3}\rho_{0}^{1/3}
\end{equation}

\noindent
so that the net energy release due to colliding matter is

\begin{equation}\label{E:EqA4}
U = \frac{\sqrt[\leftroot{2} \uproot{2} 3]{36\pi} }{5} G\rho_{0}^{1/3} M^{5/3} \left( 1-(1-f)^{5/3}-f^{5/3} \right)
\end{equation}

\noindent
for colliding bodies of mass $fM$ and $(1-f)M$. If the two colliding Giant black holes both have initial net-trace charges, the net electogravitational potential energy release would be smaller, as would the fraction of the electron-rich outer rind that would need to be impact-jetted to initiate the Big Bang's Coulomb explosion.  

Let us compare the relative size of the black hole's core of clumped matter relative to its Schwarzschild radius. For a clumped mass $M=\SI{1e51}{kg}$ and a star-like mean density $\rho_{0}=\SI{1}{Mgm^{-3}}$, the radius of the clumped mass (eqn. A3) is $\SI{6.2e15}{m} \; (\SI{0.66}{ly})$, only $\num{4.2e-9}$ the mass's Schwartzschild radius of $\SI{1.5e24}{m} \; (\SI{1.6e8}{ly})$. Initial expansion of a Coulomb explosion of the clumped mass would occur under conditions where radiated photons would remain trapped within the still-present Giant black hole.

\section{Table 1: Nomenclature}
\begin{longtable}[l]{| p{37pt} | p{98pt} | p{150pt} |}
\caption{List of Symbols and their Values/Units.} \label{tab:long} \\
\hline
\textbf{Symbol}	& \textbf{Value (Units)} & \textbf{Description} \\
\hline 

		$\boxplus$	 	& Ckg$^{-1} \approx \epsCG$	 & + trace-charged matter\\ \hline
		$\boxminus$	 	& Ckg$^{-1} \approx \epsCG$	 & -- trace-charged matter\\ \hline
		$\square$	 	         & Ckg$^{-1} << \epsCG$	         & nearly uncharged/neutral matter\\ \hline
		$\boxemptyplusCG$       & $ << \epsCG$ Ckg$^{-1}$ & charge/mass ratio of + nearly neutral matter\\ \hline 
$\boxemptyminusCG$    & $<< \epsCG$ Ckg$^{-1}$) & charge/mass ratio of -- nearly neutral matter\\ \hline 
$a$	 	         & $\SI{7.565e-16}{JK^{-4}m^{-3}}$	         & radiation density constant\\ \hline 
$A(R_{L})$	 	 & kg	         & $\tcmuboxplus$ attractive mass-equivalent of matter within galactic radius $R_{L}$\\ \hline 
$c$	 	& $\SI{2.998e8}{ms^{-1}}$		 & vacuum speed of light\\ \hline 
$\cboxplus$	 	& ---		 & charge fraction of $\tcmuboxplus$\\ \hline 
$\cboxminus$	 	& ---		 & charge fraction of $\tcmuboxminus$\\ \hline 
$\cBH,\cGBH$	 	& ---		 & charge fraction of a black hole, central galactic blackhole\\ \hline 
$\codot$	 	& ---		 & charge fraction of nearly uncharged stellar material\\ \hline 
$\csplus$	 	& $ms^{-1}$ & isothermal speed of sound in $\tcmuboxplus$  $(\csplus = \sqrt{ \rhoplus k_{B}T / (\moverbarplus m_{p})} )$\\ \hline 
$e^{-}$	 	& $\SI{1.6022e-19}{C}$	 & Elementary charge unit of an electron or proton\\ \hline 
$\epsCG$	 	& \SI{8.6167e-11}{Ckg^{-1}}	 & Critical charge/mass ratio for electrogravitational stability\\ \hline 
$\fboxdot$	 	& ---	         & volume fraction of $\tcmuboxdot$-HI clouds \\ \hline 
$f$	 	         & ---	         & fraction of excess negative charge in rind to net core positive charge\\ \hline 
$G$	 	         & $\SI{6.673e-11}{Nm^{2}kg^{2}}$	         & Gravitational constant\\ \hline 
$G_{eff}$	 	 &Nm$^{2}$kg$^{2}$	         & reduced or effective G\\ \hline 
$\gamma$	 & ---	         & Lorentz factor $\gamma \equiv 1/ \sqrt{1-v^{2}/c^{2}}$ \\ \hline 
HI	 	         &$\ce{H1}$	         & monatomic hydrogen\\ \hline 
$k_{B}$	 	         &$\SI{1.381e-23}{JK^{-1}}$	         & Boltzmann constant\\ \hline 
$k_{E}$	 	         &$\SI{8.988e+9}{Nm^{2}C^{-2}}$	         & Coulomb's constant\\ \hline 
$\Mboxplus (R)$        & kg         & radially cumulative $\tcmuboxplus$ in dark-matter Galactic halos\\ \hline 
$\Mboxdot (R)$	 	 & kg	         & radially cumulative $\tcmuboxdot$-HI in dark-matter Galactic halos\\ \hline 
$\ML (R)$	 	         & kg	         & radially cumulative  luminous matter in a Galaxy excluding $\Mboxdot (R)$\\ \hline 
$\MGBH$	 	 & kg	         & central Galactic Black Hole mass\\ \hline 
$\Msun$	 	         & \SI{1.99e30}{kg}	         & Solar mass\\ \hline 
$m$	 	         & kg	         & mass\\ \hline 
$m_{e}$	 	         & $\SI{9.109e-31}{kg}$	         & electron mass\\ \hline 
$m_{p}, m_{{}_{H^{+}}}$	 	         & $\SI{1.673e-27}{kg}$	         & proton (also called $\ce{H+}$) mass\\ \hline 
$\moverbarplus$	 	         & ---	         & mean particle mass of $\tcmuboxplus /m_{p}$\\ \hline 
$\moverbardot$	 	 	         & ---	         & mean particle mass of $\tcmuboxdot /m_{p}$\\ \hline 
$\tcmuboxminus$	 	         & ---	         & negatively trace-charged matter\\ \hline 
$\mu_{0}$	 	         & 4$\pi \times 10^{-7}$NA$^{-2}$ & magnetic constant\\ \hline 
$(\tcmuboxplusplusplus),\tcmuboxplus$	 	 	         & ---	         & (hyper-positive) and positively trace-charged matter\\ \hline 
$\tcmuboxdot$	 	                 & ---	         & quasi-neutral matter (usually bright or radioluminous) \\ \hline 
$\tcmuboxdotplus$, 	$\tcmuboxdotminus$ & --- & ($+$,$-$) quasi-neutral matter  \\ \hline 
$\tcmuCG$	 	& Ckg$^{-1}$	 & charge/mass ratio of matter\\ \hline 
$\tcmuodotplus$	 	& $\SI{7.753e-29}{Ckg^{-1}}$	 & critical solar (and stellar) charge/mass ratio\\ \hline 
$n_{0}$	 	& ---	 & particle number density\\ \hline
$P,p$	 	& \si{\pascal}	 & Pressure\\ \hline
$P_{rad}$	 	& \si{\pascal}	 & radiation pressure\\ \hline
$\Qsun$	 	& C	 & Net solar charge\\ \hline
$\Qboxplus (R)$	 	& C	 & Cumulative charge associated with $\Mboxplus (R)$\\ \hline
$r,R$	 	& \si{\meter}	 & radial distance\\ \hline
$\Rsun$	 	& \SI{7e8}{m}	 & Solar radius\\ \hline
$\RL$	 	& m (or kpc)	 & Radial limit of luminous stellar matter in a spiral galaxy\\ \hline
$R_{S}$	 	&m	 & Schwarzschild radius\\ \hline
$\rho,\rho_{0}$	 	&kgm$^{-3}$	 & density, central density of of trace-charged Bonnor-Ebert dark halo\\ \hline
$\rhoplus, \rhodot$	 	&kgm$^{-3}$	 & density of $\tcmuboxplus, \tcmuboxdot$\\ \hline
$\rho_{\scriptscriptstyle{UBH}}$	 	&kgm$^{-3}$	 & mean Schwarzschild density of an ultramassive black hole\\ \hline
$R_{S}$	 	&m	 & Schwarzschild radius\\ \hline
$T$	 	& \si{\kelvin}	 & Temperature\\ \hline
$U$	 	& J	 & Binding energy of matter clump within a Giant black hole\\ \hline
$U_{\scriptscriptstyle{BH}}^{p^{+}}$	 	& J,eV	 & electrostatic work done on protons jetting from a charged black hole\\ \hline
$v$	 	& ms$^{-1}$	 & velocity\\ \hline
$V_{C}, V_{ROT}$ & ms$^{-1}$	 & Keplerian rotation curve due to gravitational attraction of $\tcmuboxplus$\\ \hline
$w_{i}$	 	& ms$^{-1}$	 & most probable thermal velocity for particle of mass $m_{i}$\\ \hline
$w_{e},w_{p}$	 	& ms$^{-1}$	 & electron, proton escape velocity \\ \hline

\end{longtable}





\bibliographystyle{plainnat}

\bibliography{bigbangbib2} 

\begin{thebibliography}{48}
\providecommand{\natexlab}[1]{#1}
\providecommand{\url}[1]{\texttt{#1}}
\expandafter\ifx\csname urlstyle\endcsname\relax
  \providecommand{\doi}[1]{doi: #1}\else
  \providecommand{\doi}{doi: \begingroup \urlstyle{rm}\Url}\fi

\bibitem[{Abreau, P. et al.}(2010)]{abre2010}
{Abreau, P. et al.}
\newblock Update on the correlation of the highest energy cosmic rays with
  nearby extragalactic matter.
\newblock \emph{Astroparticle Physics}, 34\penalty0 (5):\penalty0 314--326,
  December 2010.
\newblock ISSN 0927-6505.

\bibitem[Bahcall(2015)]{Bahc2015}
Neta~A. Bahcall.
\newblock Dark matter universe.
\newblock \emph{Proceedings of the National Academy of Sciences}, 112\penalty0
  (40):\penalty0 12243--12245, 2015.
\newblock \doi{10.1073/pnas.1516944112}.

\bibitem[Bahcall and Kulier(2014)]{Bahc2014}
Neta~A. Bahcall and Andrea Kulier.
\newblock Tracing mass and light in the universe: where is the dark matter?
\newblock \emph{Monthly Notices of the Royal Astronomical Society},
  439\penalty0 (3):\penalty0 2505--2514, 2014.
\newblock \doi{10.1093/mnras/stu107}.

\bibitem[Begeman(1987)]{bege1987}
K.~Begeman.
\newblock \emph{HI rotation curves of spiral galaxies}.
\newblock Ph.d., Uni. Gronigen, 1987.

\bibitem[Begeman(1989)]{bege1989}
K.~Begeman.
\newblock H1 rotation curves of spiral galaxies.
\newblock \emph{Astron. Astrophys.}, 223:\penalty0 47--60, 1989.

\bibitem[{Berezinskii} et~al.(1990){Berezinskii}, {Bulanov}, {Dogiel}, and
  {Ptuskin}]{bere1990}
V.~S. {Berezinskii}, S.~V. {Bulanov}, V.~A. {Dogiel}, and V.~S. {Ptuskin},
  editors.
\newblock \emph{Astrophysics of cosmic rays}.
\newblock North -Holland, 1990.

\bibitem[Bhattacharjee and Sigl(2000)]{bhat2000}
P.~Bhattacharjee and G.~Sigl.
\newblock \emph{Origin and Propagation of Extremely High-energy Cosmic Rays}.
\newblock Physics reports. Elsevier, 2000.

\bibitem[Bignami et~al.(2003)Bignami, Caraveo, Luca, and Mereghetti]{bign2003}
G.~F. Bignami, P.~A. Caraveo, A.~De Luca, and S.~Mereghetti.
\newblock The magnetic field of an isolated neutron star from x-ray cyclotron
  absorption lines.
\newblock \emph{Nature}, 423:\penalty0 725--727, 2003.

\bibitem[Binney and Tremaine(2007)]{binn2007}
J.~Binney and S.~Tremaine.
\newblock \emph{Galactic Dynamics}.
\newblock Princeton University Press, second edition, 2007.

\bibitem[Bondi(1999)]{bond1999}
H.~Bondi.
\newblock The gravitational redshift from static spherical bodies.
\newblock \emph{Mon. Not. Royal Astron. Soc.}, 302:\penalty0 337--340, 1999.

\bibitem[Bonnor(1980)]{bonn1980}
W.~Bonnor.
\newblock Equilibrium of charged dust in general relativity.
\newblock \emph{General Relativity and Gravitation}, 12:\penalty0 453--465,
  1980.

\bibitem[Bonnor(1956)]{bonn1956}
W.~B. Bonnor.
\newblock Boyle's law and gravitational unstability.
\newblock \emph{Mon. Not. Royal Astron. Soc.}, 116:\penalty0 351--359, 1956.

\bibitem[Bonnor and Wickramasuriya(1975)]{Bonn1975}
W.~B. Bonnor and S.~B.~P. Wickramasuriya.
\newblock Are very large gravitational redshifts possible?
\newblock \emph{Monthly Notices of the Royal Astronomical Society},
  170\penalty0 (3):\penalty0 643--649, 1975.
\newblock \doi{10.1093/mnras/170.3.643}.

\bibitem[Brandt(2008)]{bran2008}
H.~Brandt.
\newblock \emph{Astrophysics Processes}.
\newblock Cambridge, 2008.

\bibitem[Camilo et~al.(2006)Camilo, Ransom, Halpern, Reynolds, Helfand,
  Zimmerman, and Sarkissian]{cami2006}
F.~Camilo, S.~M. Ransom, J.~P. Halpern, J.~Reynolds, D.~J. Helfand,
  N.~Zimmerman, and J.~Sarkissian.
\newblock Transient pulsed radio emission from a magnetar.
\newblock \emph{Nature}, 442:\penalty0 892--895, 2006.

\bibitem[Cosby and Helm(1988)]{cosb1988}
P.~C. Cosby and H.~Helm.
\newblock Experimental determination of the ${H}_{3}^{+}$ bond dissociation
  energy.
\newblock \emph{Chem. Phys. Lett.}, 152:\penalty0 71--74, 1988.

\bibitem[{Doeleman, Sheperd S. et al.}(2012)]{Doel2012}
{Doeleman, Sheperd S. et al.}
\newblock Jet-launching structure resolved near the supermassive black hole in
  m87.
\newblock \emph{Science}, 338\penalty0 (6105):\penalty0 355--358, 2012.
\newblock ISSN 0036-8075.
\newblock \doi{10.1126/science.1224768}.

\bibitem[Ebert(1955)]{eber1955}
R.~Ebert.
\newblock Uber die verdichtung von h-i-gebieten.
\newblock \emph{Zeitschrift fuer Astrophysik}, 37:\penalty0 217--232, 1955.

\bibitem[Faber and Gallagher(1979)]{fabe1979}
S.~M. Faber and J.~S. Gallagher.
\newblock masses and mass-to-light ratios of galaxies.
\newblock \emph{Annu. Rev. Astron. Astrophys.}, 17:\penalty0 135--187, 1979.

\bibitem[Ghisellini et~al.(2014)Ghisellini, Tavecchio, Maraschi, Celotti, and
  Sbarrato]{ghis2014}
G.~Ghisellini, F.~Tavecchio, L.~Maraschi, A.~Celotti, and T.~Sbarrato.
\newblock The power of relativistic jets is larger than the luminosity of their
  accretion disks.
\newblock \emph{Nature}, 515\penalty0 (7527):\penalty0 376--378, November 2014.
\newblock ISSN 0028-0836.

\bibitem[{Greisen}(1966)]{grei1966}
K.~{Greisen}.
\newblock {End to the Cosmic-Ray Spectrum?}
\newblock \emph{Physical Review Letters}, 16:\penalty0 748--750, April 1966.
\newblock \doi{10.1103/PhysRevLett.16.748}.

\bibitem[Hairer et~al.(2003)Hairer, Lubich, and Wanner]{hair2003}
E.~Hairer, C.~Lubich, and G.~Wanner.
\newblock Geometric numerical integration illustrated by the stoermer-verlet
  method.
\newblock \emph{Acta Numerica}, pages 399--450, 2003.

\bibitem[Hollweg(1970)]{holl1970}
J.~V. Hollweg.
\newblock Collisionless solar wind.
\newblock \emph{J. Geophys. Res.}, 75:\penalty0 2403--2418, 1970.

\bibitem[Jockers(1970)]{jock1970}
K.~Jockers.
\newblock Solar wind models based on exospheric theory.
\newblock \emph{Astron. Astrophys.}, 6:\penalty0 219--239, 1970.

\bibitem[Lemaire and Scherer(1971)]{lema1971}
J.~Lemaire and M.~Scherer.
\newblock Kinetic models of the solar wind.
\newblock \emph{J. Geophys. Res.}, 76:\penalty0 7479--7490, 1971.

\bibitem[Lyne and Graham-Smith(2012)]{lyne2012}
A.~Lyne and F.~Graham-Smith.
\newblock \emph{Pulsar Astronomy}, volume~48 of \emph{Cambridge Astrophysics
  Series}.
\newblock Cambridge University Press, 4th edition, 2012.

\bibitem[Majumder(1947)]{maju1947}
S.~D. Majumder.
\newblock A class of exact solutions of einstein's field equations.
\newblock \emph{Phys. Rev.}, 72:\penalty0 390--398, 1947.

\bibitem[{Maksimovic} et~al.(1997){Maksimovic}, {Pierrard}, and
  {Lemaire}]{maks1997}
M.~{Maksimovic}, V.~{Pierrard}, and J.~F. {Lemaire}.
\newblock {A kinetic model of the solar wind with Kappa distribution functions
  in the corona.}
\newblock \emph{Astron. Astrophys.}, 324:\penalty0 725, 1997.

\bibitem[McCall et~al.(1999)McCall, Geballe, Hinkle, and Oka]{mcca1999}
B.~J. McCall, T.~R. Geballe, K.~H. Hinkle, and T.~Oka.
\newblock Observations of ${H}_{3}^{+}$ in dense molecular clouds.
\newblock \emph{The Astrophysical Journal}, 522\penalty0 (1):\penalty0 338,
  1999.

\bibitem[{Meyer, E.T. et al.}(2015)]{meye2015}
{Meyer, E.T. et al.}
\newblock A kiloparsec-scale internal shock collision in the jet of a nearby
  radio galaxy.
\newblock \emph{Nature}, 521\penalty0 (7553):\penalty0 495--497, May 2015.
\newblock ISSN 0028-0836.

\bibitem[Meyer-Vernet(2007)]{meye2007}
N.~Meyer-Vernet.
\newblock \emph{Basics of the Solar Wind}.
\newblock Cambridge Atmospheric and Space Science Series. Cambridge University
  Press, 2007.
\newblock ISBN 9781139461559.

\bibitem[Oka(1992)]{oka1992}
Takeshi Oka.
\newblock The infrared spectrum of ${H}_{3}^{+}$ in laboratory and space
  plasmas.
\newblock \emph{Rev. Mod. Phys.}, 64:\penalty0 1141--1149, Oct 1992.
\newblock \doi{10.1103/RevModPhys.64.1141}.

\bibitem[Oka(2006)]{Oka2006}
Takeshi Oka.
\newblock Interstellar ${H}_{3}^{+}$.
\newblock \emph{Proceedings of the National Academy of Sciences}, 103\penalty0
  (33):\penalty0 12235--12242, 2006.
\newblock \doi{10.1073/pnas.0601242103}.

\bibitem[{Olausen} and {Kaspi}(2014)]{olau2014}
S.~A. {Olausen} and V.~M. {Kaspi}.
\newblock {The McGill Magnetar Catalog}.
\newblock \emph{The Astrophysical Journal, Supplement}, 212:\penalty0 6, May
  2014.
\newblock \doi{10.1088/0067-0049/212/1/6}.

\bibitem[{Page, M. J. et al.}(2012)]{page2012}
{Page, M. J. et al.}
\newblock The suppression of star formation by powerful active galactic nuclei.
\newblock \emph{Nature}, 485\penalty0 (7397):\penalty0 213--216, May 2012.
\newblock ISSN 0028-0836.

\bibitem[Pannekoek(1922)]{Pann1922}
A.~Pannekoek.
\newblock Ionization in stellar atmospheres.
\newblock \emph{Bull. Astron. Inst. Neth.}, 1:\penalty0 107--118, 1922.

\bibitem[Pantellini et~al.(2012)Pantellini, Landi, Zaslavsky, and
  Meyer-Vernet]{pant2012}
F.~Pantellini, S.~Landi, A.~Zaslavsky, and N.~Meyer-Vernet.
\newblock On the unconstrained expansion of a spherical plasma cloud turning
  collisionless: case of a cloud generated by a nanometre dust grain impact on
  an uncharged target in space.
\newblock \emph{plasma physics and controlled fusion}, 54:\penalty0 045005,
  2012.
\newblock \doi{10.1088/0741-3335/54/4/045005}.

\bibitem[Parker(1965)]{park1965}
E.N. Parker.
\newblock Dynamic theory of the solar wind.
\newblock \emph{Space Science Reviews}, 4:\penalty0 666--708, 1965.

\bibitem[{Pavlov} et~al.(2003){Pavlov}, {Teter}, {Kargaltsev}, and
  {Sanwal}]{pavl2003}
G.~G. {Pavlov}, M.~A. {Teter}, O.~{Kargaltsev}, and D.~{Sanwal}.
\newblock {The Variable Jet of the Vela Pulsar}.
\newblock \emph{The Astrophysical Journal}, 591:\penalty0 1157--1171, July
  2003.
\newblock \doi{10.1086/375531}.

\bibitem[Pierrard and Lazar(2010)]{Pier2010}
V.~Pierrard and M.~Lazar.
\newblock Kappa distributions: Theory and applications in space plasmas.
\newblock \emph{Solar Physics}, 267\penalty0 (1):\penalty0 153--174, 2010.
\newblock ISSN 1573-093X.
\newblock \doi{10.1007/s11207-010-9640-2}.

\bibitem[{Rienstra-Kiracofe} et~al.(2002){Rienstra-Kiracofe}, Tschumper,
  {Schaefer III}, Nandi, and Ellison]{Rien2002}
J.~C. {Rienstra-Kiracofe}, G.~S. Tschumper, H.~F. {Schaefer III}, S.~Nandi, and
  G.~B. Ellison.
\newblock Atomic and molecular electron affinities: Photoelectron experiments
  and theoretical computations.
\newblock \emph{Chem. Rev.}, 102:\penalty0 231--282, 2002.

\bibitem[Rosseland(1924)]{Ross1924}
S.~Rosseland.
\newblock Electrical state of a star.
\newblock \emph{Mon. Not. Royal Astron. Soc.}, 84:\penalty0 720--728, 1924.

\bibitem[Rubin and Ford(1970)]{rubi1970}
V.~C. Rubin and W.~K. Ford.
\newblock Rotation of the andromeda nebula from a spectroscopic survey of
  emission regions.
\newblock \emph{Astrophysical Journal}, 159:\penalty0 379--403, 1970.

\bibitem[Sofue and Rubin(2001)]{sofu2001}
Y.~Sofue and V.~Rubin.
\newblock Rotation curves of spiral galaxies.
\newblock \emph{Annu. Rev. Astron. Astrophys.}, 39:\penalty0 137--174, 2001.

\bibitem[{Weisskopf, M.C. et al.}(2000)]{weis2000}
{Weisskopf, M.C. et al.}
\newblock Discovery of spatial and spectral structure in the x-ray emission
  from the crab nebula.
\newblock \emph{The Astrophysical Journal}, 536:\penalty0 L81--L84, June 2000.

\bibitem[Weyl(1917)]{weyl1917}
H.~Weyl.
\newblock Zur gravitationstheorie.
\newblock \emph{Annalen der Physik}, 359:\penalty0 117--145, 1917.

\bibitem[{Zatsepin} and {Kuz'min}(1966)]{zats1966}
G.~T. {Zatsepin} and V.~A. {Kuz'min}.
\newblock {Upper Limit of the Spectrum of Cosmic Rays}.
\newblock \emph{Soviet Journal of Experimental and Theoretical Physics
  Letters}, 4:\penalty0 78, August 1966.

\bibitem[Zhang et~al.(2004)Zhang, Cheng, Kim, Stanojevic, and Eyler]{zhan2004}
Y.~P. Zhang, C.~H. Cheng, J.~T. Kim, J.~Stanojevic, and E.~E. Eyler.
\newblock Dissociation energies of molecular hydrogen and the hydrogen
  molecular ion.
\newblock \emph{Phys. Rev. Lett.}, 92:\penalty0 203003, May 2004.
\newblock \doi{10.1103/PhysRevLett.92.203003}.

\end{thebibliography}


%
%


\end{document}